\documentclass[conference]{IEEEtran}
\IEEEoverridecommandlockouts
\usepackage{cite}
\usepackage{amsmath,amssymb,amsfonts}
\usepackage{graphicx}
\usepackage{textcomp}
\usepackage{xcolor}
\usepackage{xspace}
\usepackage{hyperref}
\usepackage{tabu}
\usepackage{algorithm}
\usepackage[noend]{algpseudocode}
\usepackage{array}
\usepackage[T1]{fontenc}
\usepackage{enumitem}
\usepackage{multirow}
\usepackage{longtable}
\usepackage{adjustbox}

\usepackage[noend]{algpseudocode}

\newcolumntype{P}[1]{>{\centering\arraybackslash}p{#1}}
\newcolumntype{M}[1]{>{\centering\arraybackslash}m{#1}}

\usepackage{pifont}
\usepackage{xcolor}
\newcommand{\cmark}{\textcolor{green!80!black}{\ding{51}}}
\newcommand{\xmark}{\textcolor{red}{\ding{55}}}
\newcommand{\WF}{{\textsc{Stitcher}}\xspace} % Stitcher name
\newcommand{\ST}{\WF} % Stitcher name
\newcommand{\WG}{{\textsc{Scope}}\xspace} % SCOPE name
\newcommand{\SP}{\WG} % SCOPE name

\usepackage{pifont}
\usepackage{xcolor}

\def\BibTeX{{\rm B\kern-.05em{\sc i\kern-.025em b}\kern-.08em
    T\kern-.1667em\lower.7ex\hbox{E}\kern-.125emX}}

\usepackage{tikz,xcolor,hyperref}

\definecolor{lime}{HTML}{A6CE39}
\DeclareRobustCommand{\orcidicon}{%
	\begin{tikzpicture}
	\draw[lime, fill=lime] (0,0) 
	circle [radius=0.16] 
	node[white] {{\fontfamily{qag}\selectfont \tiny ID}};
	\draw[white, fill=white] (-0.0625,0.095) 
	circle [radius=0.007];
	\end{tikzpicture}
	\hspace{-2mm}
}

\foreach \x in {A, ..., Z}{%
	\expandafter\xdef\csname orcid\x\endcsname{\noexpand\href{https://orcid.org/\csname orcidauthor\x\endcsname}{\noexpand\orcidicon}}
}

\begin{document}

\title{A Smart City Infrastructure Ontology for Threats, Cybercrime, and Digital Forensic Investigation\\
}

\author{\IEEEauthorblockN{Yee Ching Tok\orcidA{}}
\IEEEauthorblockA{\textit{Singapore Univ. of Tech. and Design} \\
Singapore \\
yeeching\_tok@sutd.edu.sg}
\and
\IEEEauthorblockN{Davis Yang Zheng\orcidB{}}
\IEEEauthorblockA{\textit{Singapore Univ. of Tech. and Design} \\
Singapore \\
davis.zheng@owasp.org}
\and
\IEEEauthorblockN{Sudipta Chattopadhyay\orcidC{}}
\IEEEauthorblockA{\textit{Singapore Univ. of Tech. and Design} \\
Singapore \\
sudipta\_chattopadhyay@sutd.edu.sg}
% \and
% \and
% \and
}

\maketitle
\thispagestyle{plain}
\pagestyle{plain}

\begin{abstract}
Cybercrime and the market for cyber-related compromises are becoming attractive revenue sources for state-sponsored actors, cybercriminals and technical individuals affected by financial hardships. Due to burgeoning cybercrime on new technological frontiers, efforts have been made to assist digital forensic investigators (DFI) and law enforcement agencies (LEA) in their investigative efforts.

Forensic tool innovations and ontology developments, such as the Unified Cyber Ontology (UCO) and Cyber-investigation Analysis Standard Expression (CASE), have been proposed to assist DFI and LEA. Although these tools and ontologies are useful, they lack extensive information sharing and tool interoperability features, and the ontologies lack the latest Smart City Infrastructure (SCI) context that was proposed.

To mitigate the weaknesses in both solutions and to ensure a safer cyber-physical environment for all, we propose the Smart City Ontological Paradigm Expression (\SP), an expansion profile of the UCO and CASE ontology that implements SCI threat models, SCI digital forensic evidence, attack techniques, patterns and classifications from MITRE.

We showcase how \SP could present complex data such as SCI-specific threats, cybercrime, investigation data and incident handling workflows via an incident scenario modelled after publicly reported real-world incidents attributed to Advanced Persistent Threat (APT) groups. We also make \SP available to the community so that threats, digital evidence and cybercrime in emerging trends such as SCI can be identified, represented, and shared collaboratively.
\end{abstract}

% \begin{IEEEkeywords}
% component, formatting, style, styling, insert
% \end{IEEEkeywords}

\section{Introduction} 
\label{Introduction}

Cybercrime and the market for cyber-related compromises are becoming attractive revenue sources for state-sponsored actors, cybercriminals and technical individuals affected by financial hardships. The expected economic impact of cybercrime will reach US\$10.5 trillion by 2025~\cite{CyberSecVentures_2020}, making it an appealing choice compared to riskier ventures such as drug trafficking and piracy. Cybercrime and cyber threats were also critical concerns on the international front, ranking 8th out of 32 global risks in severity from short and long-term perspectives~\cite{WorldEconomicForum_2023}. 

Due to burgeoning cybercrime on new technological frontiers, efforts have been made to assist digital forensic investigators (DFI) and law enforcement agencies (LEA) in their investigative efforts. Tools such as \ST~\cite{TOK_2020301071} and smart city infrastructure (SCI) related threat models~\cite{TOK_2023} have been proposed as possible solutions to correlate evidence, identify threats and aid in digital forensic investigations in these nascent areas. Meanwhile, in a bid to enhance information sharing and tool interoperability, ontologies such as the Unified Cyber Ontology (UCO)~\cite{syed2016uco} and the Cyber-investigation Analysis Standard Expression (CASE)~\cite{CASEY201714} were suggested. Although the previously mentioned tools and ontologies are helpful, some inadequacies exist. The tools lacked extensive information sharing and tool interoperability features, while the ontologies lacked the latest SCI context that was proposed~\cite{TOK_2023}.

To mitigate the weaknesses in both solutions highlighted previously and to ensure a safer cyber-physical environment for all, we propose the Smart City Ontological Paradigm Expression (\SP), an expansion profile of the UCO and CASE ontology that implements the research information obtained from the SCI threat model~\cite{TOK_2023}. As a start, various cyber threats are categorized under Spoofing, Tampering, Repudiation, Information Disclosure, Denial of Service and Elevation of Privilege (STRIDE). The SCI data structures within \SP are technology-agnostic while adhering to international standards such as ISO37101:2016, ISO37120:2018, ISO37122:2019 and ISO37123:2019~\cite{ISO_37101_2016, ISO_37120_2018, ISO_37122_2019, ISO_37123_2019}. Finally, digital forensic evidence related to SCI threats identified via threat modelling is also included in \SP.

The contributions of our research are summarized as follows:
\begin{enumerate}
    \item We observed the data model used in UCO and CASE, and proposed a SCI-focused ontological profile known as \SP. Following the goals of UCO and CASE, \SP supports coordinated cyber investigations and tool interoperability. Furthermore, \SP also aligns with the cybersecurity industry and practitioners by integrating attack techniques, attack patterns and classifications from MITRE.
    
	\item We showcase how \SP could be used to present complex data such as SCI-specific threats, cybercrime, investigation data, and incident handling workflows via an incident scenario modelled after publicly reported real-world incidents attributed to Advanced Persistent Threat (APT) groups. We also further compare SCI representations using \SP and UCO/CASE, showcasing the ease of usage if \SP is adopted into UCO and CASE. 
 
	\item We make available \SP to the community so that threats, digital evidence and cybercrime in emerging trends such as SCI can be identified, represented, and shared in a collaborative manner. 
\end{enumerate}

For reproducibility and advancing the research in SCI, UCO and CASE, \SP and its associated contents are publicly available at: \url{https://github.com/scopeProject}{.}

The rest of this paper is organized as follows. In Section~\ref{Background}, we present the background and motivation of the paper. In Section~\ref{SCOPE}, we showcase the structure of SCOPE and the various examples when \SP is used for digital forensic investigations. In Section~\ref{SCOPE_Evaluation}, we evaluate \SP and showcase its differences from existing ontologies. In Section~\ref{Limitations}, we highlight the limitations of our research and \SP. In Section~\ref{RelatedWork}, we summarize current related work. Finally, we conclude the paper in Section~\ref{Conclusion}. 
\section{Background and Motivation} 
\label{Background}

Digital forensic investigations require not only technical training 
but also background knowledge related to the investigated platforms. This is to complete their investigations effectively. In a user study conducted by Tok et al.~\cite{TOK_2020301071}, it was observed that DFI who had undergone professional training or worked on prior hands-on investigations performed better when faced with the given domain-specific forensic scenario. However, SCI is an emerging area with multiple forms of interpretation of the area~\cite{TOK_2023}. Freshly graduated DFI from Institutes of Higher Learning (IHL) may also need help in SCI investigation as the breadth and depth of teaching and assessments vary between various IHLs. This observation is further supported by the user study conducted by Tok et al.~\cite{TOK_2020301071}, where more than half of the participants that studied digital forensics in an IHL {\em failed to solve the given forensic scenario}. This presents a worrying scenario - DFI \textit{might} be overwhelmed when attacks on SCI become prevalent, as they already face multiple challenges in their investigation work in traditional domains~\cite{TOK_2020301071}.

\subsection{Sharing Forensic Data in Emerging Technologies} \label{Background_DataSharing}

Other than digital forensic data, DFI and LEA also have to contend with different types of data, such as digital threats and types of cybercrime committed. Traditional commercial digital forensic tools such as EnCase~\cite{EnCase2} and FTK~\cite{AccessData2} dealt solely with data management and forensic processes on conventional platforms such as desktop computers, laptops and mobile devices. However, these tools do not include other data that would provide context. Current electronic discovery (e-Discovery) tools only work on traditional platforms, but no support for SCI-related platforms is mentioned~\cite{TOK_2023}.

Digital investigations may also involve multiple parties and evidence formats, especially in the case of complex systems such as SCI. As such, collaborative investigations, data-sharing and tool updates/innovations are inevitable requirements. File formats such as the Advanced Forensic Format (AFF4) could be used by tools to store some digital forensic metadata via the Resource Description Framework (RDF)~\cite{COHEN2009S57}. Information-sharing frameworks such as Structured Threat Information Expression (STIX) and Trusted Automated Exchange of Intelligence Information (TAXII) have also been proposed but were focused more on cyber threat intelligence~\cite{STIXTAXII_2023}. Meanwhile, digital forensic innovations and tool updates may take time. For example, a widely used tool in digital forensics, \textit{bulk\_extractor}, took longer than anticipated for a subsequent update~\cite{Garfinkel_CACM_2023}. However, the end result yielded impressive performance optimizations and significant improvements in code quality and reliability~\cite{Garfinkel_CACM_2023}.

\subsection{Representing Forensic Data in Emerging Technologies} \label{Background_DataRepresentation}
There is a clear technical gap for representing and sharing data, particularly threats, cybercrime, and digital forensic information for emerging technologies such as SCI platforms. Some solutions have been proposed, such as the Unified Cyber Ontology (UCO) and the Cyber-investigation Analysis Standard Expression (CASE) ontology. UCO and CASE could be used to share threats, cybercrime and digital forensic information. A few example implementations for popular digital forensic tools such as Cellebrite, Magnet Forensics and MSAB XRY have been produced~\cite{CASE-Implementation-UFED-XML_2024, CASE-Implementation-Magnet-Axiom_2024, CASE-Implementation-MSAB-XRY_2024}. However, UCO and CASE ontologies are not geared towards SCI. In fact, CASE is an extension of UCO, which includes investigation-specific ontologies. Noting the challenges DFI are grappling with regarding investigation workloads and background knowledge, it would be hard to expect DFI to adapt UCO and CASE immediately to an SCI context. Therefore, we decided to build on our previous research data~\cite{TOK_2023} and create a modular expansion that centers around SCI named Smart City Ontological Paradigm Expression (\SP). By creating this expansion, we offload the time-consuming process of creating an SCI investigation framework and associated SCI evidence artifact representation. DFI can spend more time investigating and clearing cases instead of grappling with uncertain variables such as ad-hoc forensic data representation in emerging technologies.

\subsection{Supporting Cyber Defense in Emerging Technology Deployments} \label{Background_SupportAdvancedActivities}

As emerging technologies and hardware are consolidated and integrated into SCI deployments on a city-to-national level, best practices such as product certifications and security testing would be expected. However, such certifications and security testing are merely product-specific and do not account for scenarios where various hardware components, protocols and frameworks are integrated on a city-to-national level. Governments looking to implement SCI would be concerned about the resiliency of SCI against cyberattacks from various adversaries (such as cybercriminals and APTs). Advanced security assessment activities such as adversary emulation~\cite{ATTCK_MITRE_AE_2024} would be an excellent way to assess SCI deployments comprehensively. Such activities are driven by attack techniques cataloged by organizations (e.g., MITRE) and acknowledged by the cybersecurity industry. To facilitate cybercrime investigation on a city-to-national level, \SP has fully integrated the MITRE ATT\&CK framework~\cite{ATTCK_MITRE_2024} and MITRE Common Attack Pattern Enumerations and Classifications (CAPEC)~\cite{CAPEC_MITRE_2024}. By doing so, investigators and defenders can instantly map attack techniques and patterns utilized by adversaries and associated threats/digital forensic evidence during their investigation. Coupled with the ability to share attacks and digital forensic evidence, \SP provides cyber defenders opportunities to enhance detection and response capabilities as information recorded via \SP could be utilized to improve policies, procedures and technologies while adversary emulation activities are ongoing.
% (3rd section)

\section{Smart City Ontological Paradigm Expression} \label{SCOPE}

The Smart City Ontological Paradigm Expression (\SP) extends the prior work of UCO and CASE, and is specified via Web Ontology Language OWL 2 (following the documentation from CASE~\cite{CASE-Style-Guide-2024}). UCO was designed to support information representation across multiple cyber domains, while CASE was an investigation-specific module extended from UCO~\cite{CASEY201714}. \SP aims to provide a SCI-specific extension that can be used by any interested users while maintaining the collaborative and interoperability nature of UCO/CASE. 

\subsection{Motivations for the Creation of \SP}
Initially, we envisioned using UCO/CASE based on its ontological structure and expanding it slightly for SCI-related digital forensic activities. However, it became evident that it would be challenging as smart city-specific terminologies, attack techniques, and patterns were absent within UCO/CASE. That meant extensive modification and expansion would be required if DFI adopted the usage of UCO/CASE. A one-off effort was possible, but it had its drawbacks and limitations. DFI need to know SCI intimately, along with associated threats, attack techniques, sources of evidence and related forensic artifacts. Unfortunately, based on prior research findings~\cite{TOK_2020301071, TOK_2023}, we have established that DFI are hard-pressed for time and unlikely to be able to spend time juggling casework and learning new knowledge concurrently.

\begin{table}[htbp]
\begin{center}
\centering
\caption{Comparison of Features Between UCO, CASE and \SP}
\resizebox{\columnwidth}{!}
% \resizebox{\textwidth}{!}
{\begin{tabu}to 0.9 \columnwidth{| M{3.8cm} | m{1cm} | m{1cm} |m{1cm}|} 
    \cline{1-4}
    % \multicolumn{1}{|c|}
    \textbf{Features}  & \vfil\hfil\textbf{UCO} & \vfil\hfil\textbf{CASE} & \vfil\hfil\textbf{\SP}  \\
    \hline
    \vspace*{3pt}Extensible\vspace*{3pt} & \vfil\hfil\cmark & \vfil\hfil\cmark & \vfil\hfil\cmark \\
    \hline
    \vspace*{3pt}Information Representation for Cyber Domains\vspace*{3pt}  & \vfil\hfil\cmark & \vfil\hfil\cmark & \vfil\hfil\cmark \\
    \hline
    \vspace*{3pt}Supports Cyber Investigations\vspace*{3pt} & \vfil\hfil\xmark & \vfil\hfil\cmark & \vfil\hfil\cmark \\
    \hline
    \vspace*{3pt}Smart City Infrastructure Terminology Representation\vspace*{3pt} & \vfil\hfil\xmark & \vfil\hfil\xmark & \vfil\hfil\cmark \\
    \hline    
    \vspace*{3pt}Smart City Infrastructure Threats Representation\vspace*{3pt} & \vfil\hfil\xmark & \vfil\hfil\xmark & \vfil\hfil\cmark \\
    \hline    
    \vspace*{3pt}Smart City Infrastructure Cybercrime Classifications\vspace*{3pt} & \vfil\hfil\xmark & \vfil\hfil\xmark & \vfil\hfil\cmark \\
    \hline    
    \vspace*{3pt}Smart City Infrastructure Data Sources\vspace*{3pt} & \vfil\hfil\xmark & \vfil\hfil\xmark & \vfil\hfil\cmark \\
    \hline    
    \vspace*{3pt}Smart City Infrastructure Digital Evidence Sources\vspace*{3pt} & \vfil\hfil\xmark & \vfil\hfil\xmark & \vfil\hfil\cmark \\
    \hline    
    \vspace*{3pt}MITRE ATT\&CK Framework Representation\vspace*{3pt} & \vfil\hfil\xmark & \vfil\hfil\xmark & \vfil\hfil\cmark \\
    \hline    
    \vspace*{3pt}MITRE CAPEC Framework Representation\vspace*{3pt} & \vfil\hfil\xmark & \vfil\hfil\xmark & \vfil\hfil\cmark \\
    \hline
\end{tabu}
}
\label{3_table:1}
\end{center}
\end{table}

To mitigate the impending problem, we concluded that a modular extension that abides by the ontological design of UCO/CASE but contains all the necessary information for SCI terminologies, attack techniques, patterns and forensic artifacts should be created. Table~\ref{3_table:1} shows the additional features offered by \SP by extending the prior work of UCO and CASE. By creating \SP, we offer DFI tasked with investigating SCI cybercrime substantial time savings (from repetitive literature review, research and conflicting opinions) and a comprehensive framework encompassing cyber threats, crimes, attack techniques and digital evidence.

When we explored UCO/CASE, we noticed some description entries utilized sources such as Wikipedia for some of the technical descriptions~\cite{UCO-1.3.0-docs_2024, CASE-1.3.0-docs_2024}. \SP avoids this and pulls references from established sources such as ISO standards which are not as easily mutable to illegitimate changes. \SP also extends terminologies to cover emerging technology as an effort to reduce technological obsolescence. 

\subsection{Overview of \SP} \label{Chapt5_SCOPE_Overview}
\begin{table} [htbp]
\caption{\SP Ontology Summary}
\centering
\begin{tabu} to 0.9 \columnwidth {|m{1.5cm}|m{3.4cm}|m{2.5cm}|}  
    \cline{1-3}
    % \multicolumn{1}{|c|}
    \vspace*{6pt}\textbf{\SP v0.1.0} \vspace*{6pt} & \vspace*{6pt} \hfil\textbf{Description} \vspace*{6pt}& \hfil\textbf{Justification}   \\
    \hline
    \vspace*{6pt}scope-crime\vspace*{6pt} & \vspace*{6pt} An action which constitutes an offense and is punishable by law (in SCI context). \vspace*{6pt} & \vspace*{6pt} A common set of SCI-related cybercrime are listed here. \vspace*{6pt}  \\ \cline{1-3}
    \vspace*{6pt}scope-evidence\vspace*{6pt} & \vspace*{6pt} Additional observables (evidence in our case) that is present in SCI and Internet-of-Things (IoT) devices that are currently not listed in uco-observable are placed here. \vspace*{6pt} & \vspace*{6pt} This ontology follows the definition of Smart City Infrastructure (SCI) presented in this paper~\cite{TOK_2023}. \vspace*{6pt}  \\ 
    \cline{1-3}
    \vspace*{6pt}scope-indicators\vspace*{6pt} & \vspace*{6pt}  These are the data indicators used in SCI. For ease of reference, the indicators are specified by their respective ISO clause number and comments specifying the details of the indicator. \vspace*{6pt} & \vspace*{6pt} This ontology defines the indicators used in Smart City Infrastructure (SCI). For a more detailed explanation on how the indicators are grouped, please refer to the paper~\cite{TOK_2023}. \vspace*{6pt}  \\ 
    \cline{1-3}
    \vspace*{6pt}scope-infrastructure\vspace*{6pt} & \vspace*{6pt} This ontology defines infrastructure within a Smart City Infrastructure (SCI). The definition of SCI defined in this paper~\cite{TOK_2023} is preferred, although additional infrastructure types can be added in future to suit deployment needs or as SCI gets more mature. \vspace*{6pt} & \vspace*{6pt} A technology-agnostic definition of SCI is provided. \vspace*{6pt}  \\ 
    \cline{1-3}
    \vspace*{6pt}scope-role\vspace*{6pt} & \vspace*{6pt} This ontology defines roles present in a Smart City Infrastructure (SCI) cybercrime investigation, identification of digital forensic opportunities and threat modeling. \vspace*{6pt} & \vspace*{6pt} These roles do not exist in UCO/CASE. \vspace*{6pt}  \\ 
    \cline{1-3}
    \vspace*{6pt}scope-threats\vspace*{6pt} & \vspace*{6pt} This ontology defines threats that are identified within Smart City Infrastructure (SCI). \vspace*{6pt} & \vspace*{6pt} These SCI-specific threats are not listed inside UCO/CASE. \vspace*{6pt}  \\ 
    \cline{1-3}
    \vspace*{6pt}scope-vocabulary\vspace*{6pt} & \vspace*{6pt} Vocabularies used in Smart City Infrastructure (SCI) \vspace*{6pt} & \vspace*{6pt} These SCI-related vocabulary and attacks are not listed inside UCO/CASE. \vspace*{6pt}  \\ 
    \cline{1-3}
    \end{tabu}
    \label{5_table:1}
\end{table}

We provide an expanded ontology on SCI-focused cybercrime, digital evidence, data indicators, infrastructure, roles and threats. Users could leverage \SP to represent threats, cybercrime, and digital forensic information related to SCI. Table~\ref{5_table:1}  shows the key elements of \SP Ontology. 

\subsection{\SP Developments} 
\label{SCOPE_Developments}
Since the preliminary version of \SP, as illustrated in the preceding section, 
%~\cite{TOK_Dissertation_2023}, 
additional amendments and extensions were incorporated during the iterative process of improving and scenario-based testing. 
Such an iterative process was followed as a natural step to %cover 
provide coverage over any areas that were found lacking. 
%As such, the current version of \SP is at version 0.1.1 and 
Table~\ref{5_table:2} shows the added extensions 
to \SP after further refining our work.

% \todo{update figures and replace listing with pictures instead.}

\begin{table} [htbp]
\caption{\SP Ontology Extensions}
\centering
\begin{tabu} to 0.9 \columnwidth {|m{1.5cm}|m{3.4cm}|m{2.5cm}|}  
    \cline{1-3}
    % \multicolumn{1}{|c|}
    \vspace*{6pt}\textbf{\SP v0.1.1} \vspace*{6pt} & \vspace*{6pt} \hfil\textbf{Description} \vspace*{6pt}& \hfil\textbf{Extensions}   \\
    \hline
    \vspace*{6pt}scope-attackpatterns\vspace*{6pt} & \vspace*{6pt} The Common Attack Pattern Enumeration and Classification (CAPEC) from MITRE was integrated to provide a publicly available catalog of common attack patterns that helps users understand how adversaries exploit weaknesses in applications and other cyber-enabled capabilities.  \vspace*{6pt} & \vspace*{6pt}New entries that defines the attack patterns used in Smart City Infrastructure (SCI). \vspace*{6pt}  \\ \cline{1-3}
    \vspace*{6pt}scope-indicators\vspace*{6pt} & \vspace*{6pt}  Data Indicators used in SCI. Referencing from ISO standards as well as additional definitions from relevant government bodies.\vspace*{6pt} & \vspace*{6pt} Additional entries were further added to provide minute distinctions between the various indicators used. \vspace*{6pt}  \\ 
    \cline{1-3}
    \end{tabu}
    \label{5_table:2}
\end{table}

\begin{figure}[H]
\includegraphics[width=\columnwidth]{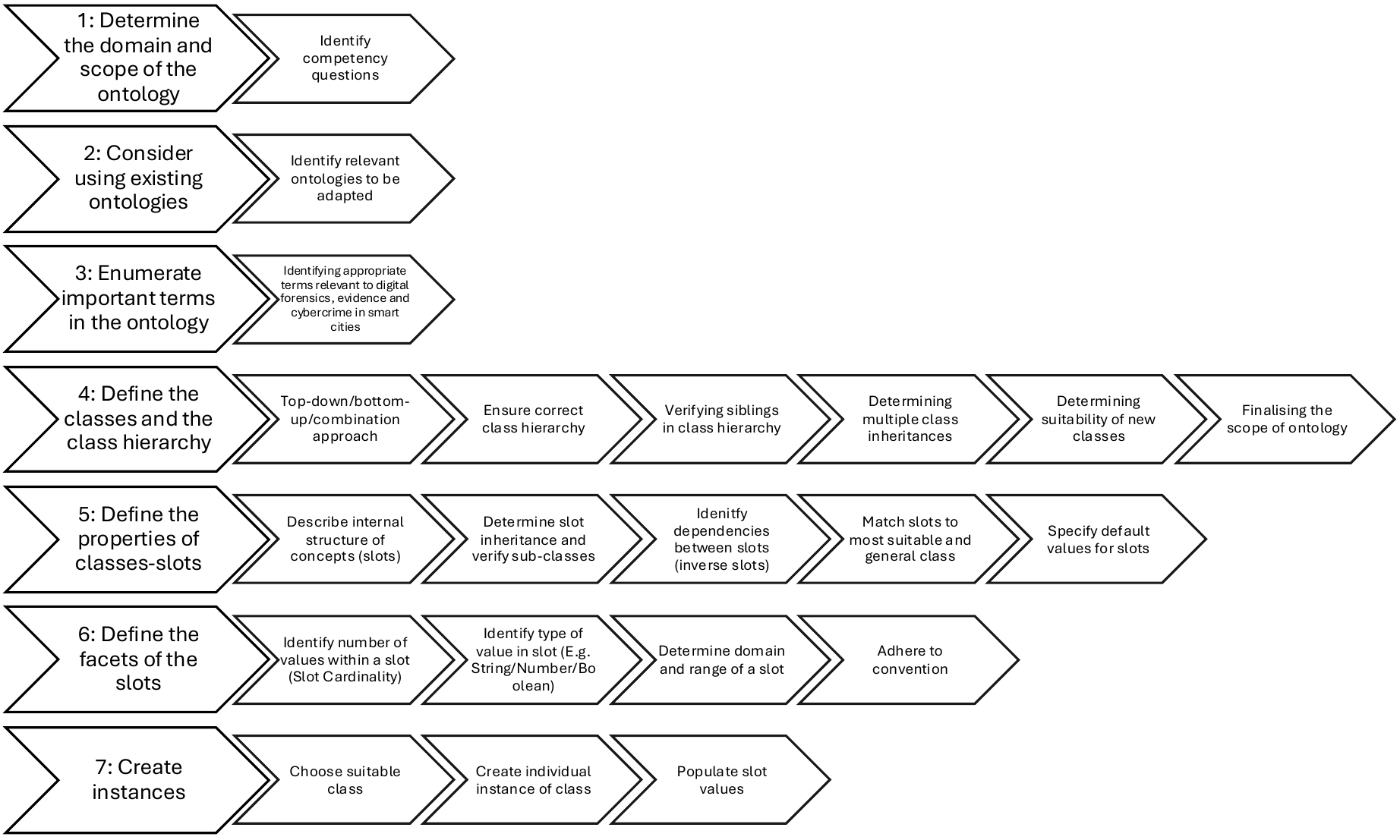}
\centering
\caption{\SP Ontology Design Steps (adapted from~\cite{Noy_2001})}
\label{fig:Ontology-Design-Steps}
\end{figure}

During the conceptualization and further optimization of \SP, we ensure that the proposed ontology is developed logically and is universally acceptable. Based on our extensive literature review of ontology design principles, we followed the recommended steps and considerations outlined in prior work~\cite{Noy_2001}. To further ensure the rigor of \SP, two researchers examined the design principles and steps outlined in the prior work of Noy and Mcguinness~\cite{Noy_2001} and ensured that \SP (via mutual checks and consensus) was developed following the considerations. A graphical representation of the recommended steps and considerations is summarized and shown in~\autoref{fig:Ontology-Design-Steps}.

In steps 1 and 2 (with reference to~\autoref{fig:Ontology-Design-Steps}), the primary objectives were to determine the domain and scope of the ontology, and also to consider using existing ontologies. The original structure of UCO/CASE was certainly robust and allowed further extensions, but a great deal of effort would be expanded to account for artifacts related to smart cities. As such, we determined an extension ontology (\SP) was the right way forward to benefit forensic investigators. In the next section, we discuss how the next design steps of \SP ontology were systematically followed.

\subsection{SCOPE Construction and Protégé}\label{Construction}

In the development of \SP, we have employed Protégé, an open-source ontology editor and a framework for building intelligent systems~\cite{Protege}. We have chosen to use Protégé as 
it provides numerous benefits while building \SP.

First and foremost, Protégé delivers a comprehensive tool which allows for the easy development and visualization of numerous 
complex relationships between the various facets of \SP. This capacity enables us to derive a more complete and accurate representation from the complex world of SCI-based digital 
forensics, and facilitates steps 3 and 4 (with reference to~\autoref{fig:Ontology-Design-Steps}) of the ontology design steps. 

Secondly, as Protégé supports the OWL~2 Web Ontology Language~\cite{OWL2_W3C_2012} and RDF specifications from the World Wide Web Consortium (W3C) such as Resource Description Framework Schema (RDFS)~\cite{RDFS_W3C_2014}, Web Ontology Language (OWL)~\cite{OWL_W3C_2012}, Semantic Web Rule Language (SWRL)~\cite{SWRL_W3C_2004} and Turtle (TTL)~\cite{Turtle_W3C_2014}. The wide range of support ensures the compatibility and interoperability 
of \SP with other related ontologies such as UCO 
and CASE, thus enhancing the applicability of \SP. 

Thirdly, the collaborative features provided by Protégé enable participation from all key stakeholders, allowing a participatory development approach (with reference to~\autoref{fig:Ontology-Design-Steps}, this would be steps 5 to 7). This enables experts from diverse backgrounds to join in the iterative development process, allowing them to lend their expertise to \SP's development, increasing its flexibility for various use cases. 

\begin{figure}[htbp]
    \centering
    \includegraphics[width=\columnwidth]{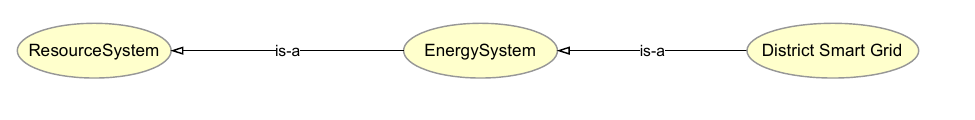}
    \caption{\SP ABox Reasoning Example}
    \label{fig:SCOPE-Abox-Reasoning}
\end{figure}

Fourthly, based on the guidelines suggested by Gruber~\cite{GRUBER1993199}, the vocabulary and ontological commitments of \SP were validated, and \autoref{fig:SCOPE-Abox-Reasoning} shows a sample of ABox reasoning of \SP. Furthermore, two researchers performed the verification work individually and reached a consensus for the final output of \SP.

Last but not least, Protégé's extensible and open-source nature provides an accessible plug-and-play environment. This makes it a flexible base for rapid prototyping and application development, allowing for customized extensions that cater to the specific needs of \SP development, including but not limited to specialized visualization tools. We leveraged Protégé to visualize the proposed \SP ontology, annotate various SCI components, threats and cybercrime, and built the conceptual map (shown in~\autoref{fig:SCOPE-Conceptual-Map}). We further used \SP to annotate the evaluation scenarios used in Section~\ref{SCOPE_Evaluation}, thus testing out the ABox reasoning for \SP. Additionally, as SCI matures, amendments and updating of \SP will be required due to the evolution of threats, cybercrime and digital evidence. Using both \SP and Protégé, users can leverage the synergy to construct an updated knowledge base that can assist with ABox reasoning. Such knowledge bases could accelerate cybercrime investigation by providing context to new investigators and insight into SCI's complexities.

In summary, the benefits mentioned in the preceding paragraph underpin the rationale of using Protégé, contributing to creating a scalable, interoperable, and comprehensive ontology that supports coordinated cyber-investigations in smart city infrastructure.

\subsubsection{Conceptual Ontology Map}
\begin{figure}[H]
    \centering
    \includegraphics[width=\columnwidth]{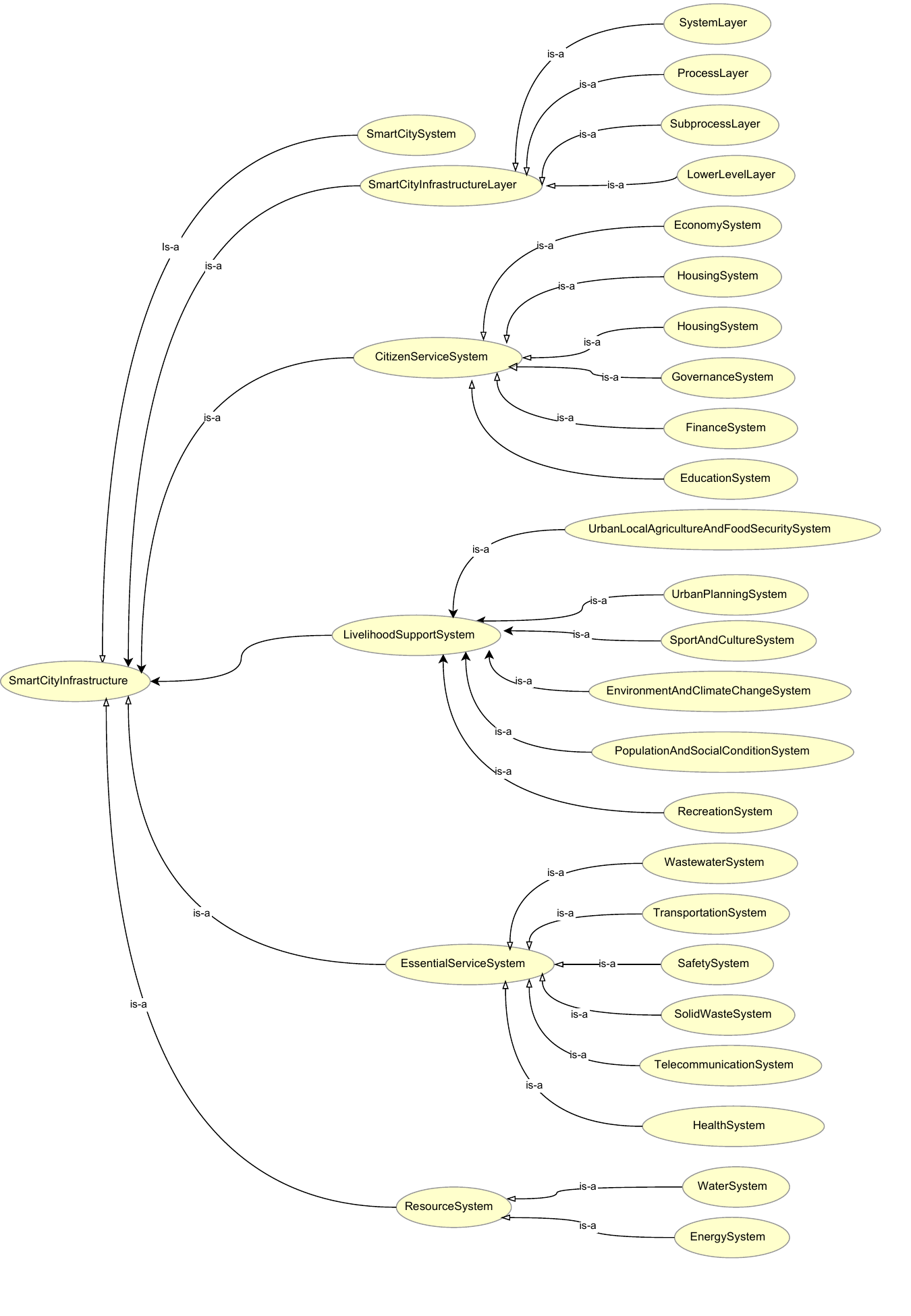}
    \caption{\SP Smart City Conceptual Map}
    \label{fig:SCOPE-Conceptual-Map}
\end{figure}

Using Protégé and with reference to~\autoref{fig:SCOPE-Conceptual-Map}, we developed a conceptual map to provide a clear understanding and overview of the structure of \SP. This map visually represents the core classes, object properties, and relationships, highlighting the hierarchical organization and interdependencies among key concepts such as threats, vulnerabilities, and assets. By illustrating these elements and their interactions, the map serves to clarify the ontology’s design, allowing for a more accessible and intuitive grasp of its complex structure and application within digital forensics and cybersecurity research.

\subsubsection{Descriptive Statistics of Ontological Concepts}
To provide a comprehensive overview of \SP’s structural elements, we include key descriptive statistics that detail its complexity. Table~\ref{tab:ontology_metrics} shows a brief summary of relevant ontological information about \SP. The ontology comprises 1,383 classes, each representing critical concepts in cybersecurity, with 180 object properties defining relational dynamics between these classes. Additionally, the ontology includes 587 data properties, enriching the model with specific characteristics for each class and 1,417 individuals, which instantiates specific entities within the ontology.

\renewcommand{\arraystretch}{0.6} % Adjust row spacing for better readability

\begin{table} [hbtp]
\caption{\SP Ontology Metrics Table}
\centering
\begin{tabular} {|p{4cm}|p{2cm}|}
 \hline
\multicolumn{2}{|c|}{\textbf{Metrics}}                    \\ \hline
\vspace*{1pt}Axiom\vspace*{1pt} & \vspace*{1pt}26,621                 \\ \hline
\vspace*{1pt}Logical axiom count\vspace*{1pt}              & \vspace*{1pt}3,748                  \\ \hline
\vspace*{1pt}Declaration axioms count\vspace*{1pt}         & \vspace*{1pt}2,200                  \\ \hline
\vspace*{1pt}Class count \vspace*{1pt}                     & \vspace*{1pt}1,383                  \\ \hline
\vspace*{1pt}Object property count \vspace*{1pt}           & \vspace*{1pt}180                    \\ \hline
\vspace*{1pt}Data property count \vspace*{1pt}             & \vspace*{1pt}587                    \\ \hline
\vspace*{1pt}Individual count\vspace*{1pt}                 & \vspace*{1pt}1,417                  \\ \hline
\multicolumn{2}{|c|}{\textbf{Class Axioms}}               \\ \hline
\vspace*{1pt}SubClassOf \vspace*{1pt}                      & \vspace*{1pt}1,462                  \\ \hline
\vspace*{1pt}DisjointClasses \vspace*{1pt}                 & \vspace*{1pt}3                      \\ \hline
\multicolumn{2}{|c|}{\textbf{Object Property Axioms}}     \\ \hline
\vspace*{1pt}SubObjectPropertyOf  \vspace*{1pt}            & \vspace*{1pt}5                      \\ \hline
\vspace*{1pt}InverseObjectProperties  \vspace*{1pt}        & \vspace*{1pt}2                      \\ \hline
\vspace*{1pt}InverseFunctionalObjectProperty\vspace*{1pt}  & \vspace*{1pt}1                      \\ \hline
\vspace*{1pt}TransitiveObjectProperty\vspace*{1pt}         & \vspace*{1pt}2                      \\ \hline
\vspace*{1pt}IrrefexiveObjectProperty\vspace*{1pt}         & \vspace*{1pt}1                      \\ \hline
\vspace*{1pt}ObjectPropertyDomain \vspace*{1pt}            & \vspace*{1pt}6                      \\ \hline
\vspace*{1pt}ObjectPropertyRange \vspace*{1pt}             & \vspace*{1pt}177                    \\ \hline
\multicolumn{2}{|c|}{\textbf{Data Property Axioms}}       \\ \hline
\vspace*{1pt}DataPropertyRange\vspace*{1pt}                & \vspace*{1pt}586                    \\ \hline
\multicolumn{2}{|c|}{\textbf{Individual Axioms}}          \\ \hline
\vspace*{1pt}ClassAssertion \vspace*{1pt}                  & \vspace*{1pt}1,447                  \\ \hline
\multicolumn{2}{|c|}{\textbf{Annotation Axioms}}          \\ \hline
\vspace*{1pt}AnnotationAssertion\vspace*{1pt}              & \vspace*{1pt}15,260                 \\ \hline
\end{tabular}
\label{tab:ontology_metrics}
\end{table}

Regarding logical structure, the ontology contains 1,462 subclass axioms and three disjoint classes, ensuring logical consistency and explicit boundaries between mutually exclusive categories. Furthermore, the ontology includes extensive annotation properties, with 15,260 annotation assertions that provide supplementary metadata, including descriptions, labels, and sources for improved interpretability. The detailed SCOPE ontology is publicly available in our website: \url{https://ontology.scopeontology.org}.

\subsection{\SP Examples} 
\label{SCOPE_Examples}

\begin{figure}[H]
\includegraphics[width=\columnwidth]{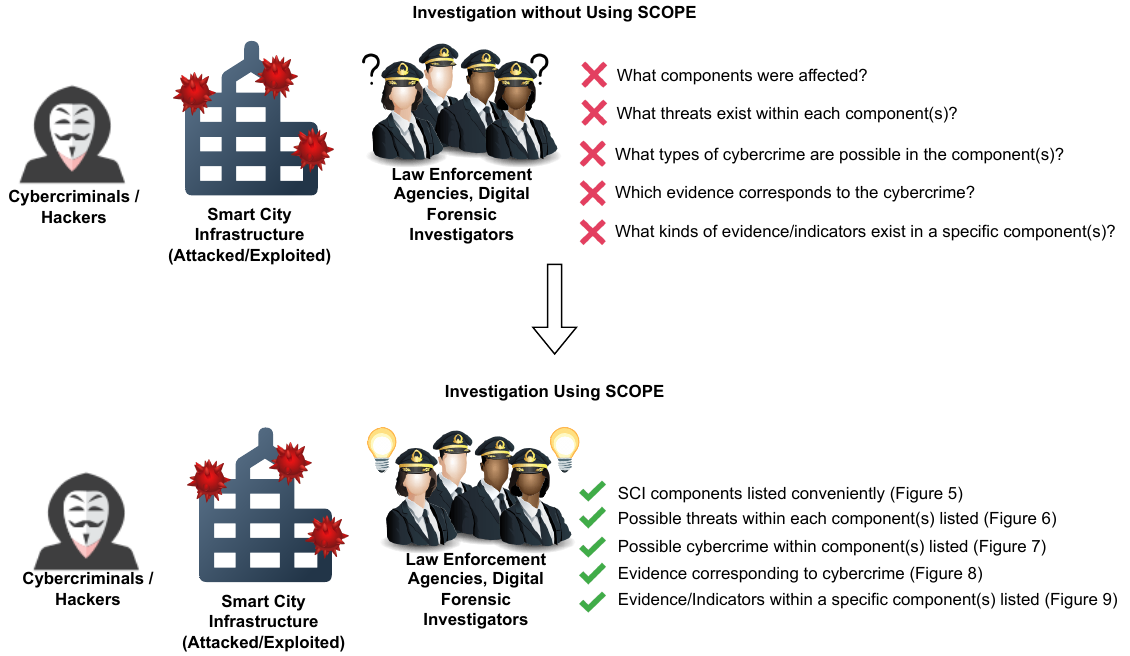}
\centering
\caption{Overview of Investigation With and Without Using \SP}
\label{fig:SCOPE-Usage-by-DFI}
\end{figure}

In this section, we showcase some theoretical examples of how \SP can be used for SCI-related threats and investigations.~\autoref{fig:SCOPE-Usage-by-DFI} shows an overview of the status quo and lists out the additional capabilities \SP can offer to DFI. For readability and illustrative purposes, the universally unique identifier (UUID) for each object is referenced with the associated system/object name (e.g., resource-system-uuid). We also import the UCO and CASE ontologies for use within our \SP ontology where needed.

\begin{figure}[htbp]
\includegraphics[width=\columnwidth]{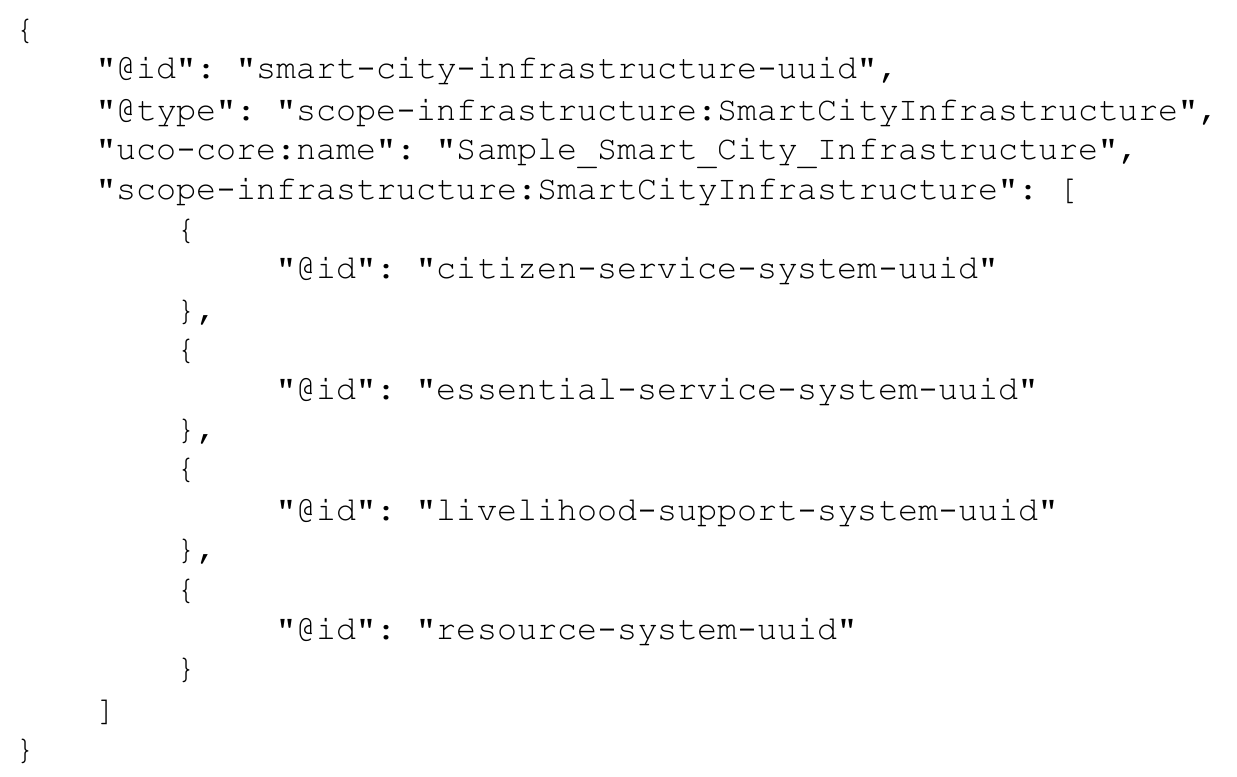}
\centering
\caption{Example of SCI in \SP Format}
\label{fig:SCOPE-SCI}
\end{figure}

With reference to~\autoref{fig:SCOPE-SCI}, we showcase an example of SCI represented in \SP format. The system layer (sometimes known as the context layer) components of a suggested SCI~\cite{TOK_2023} are outlined for ease of future use. The identifiers can be used to tag forensic evidence to the corresponding systems affected by cybercrime or map discovered attack techniques on the component.

% \todo{add 1-2 sentences to explain the above format.} - done (25/07/2024)

\begin{figure}[htbp]
\includegraphics[width=\columnwidth]{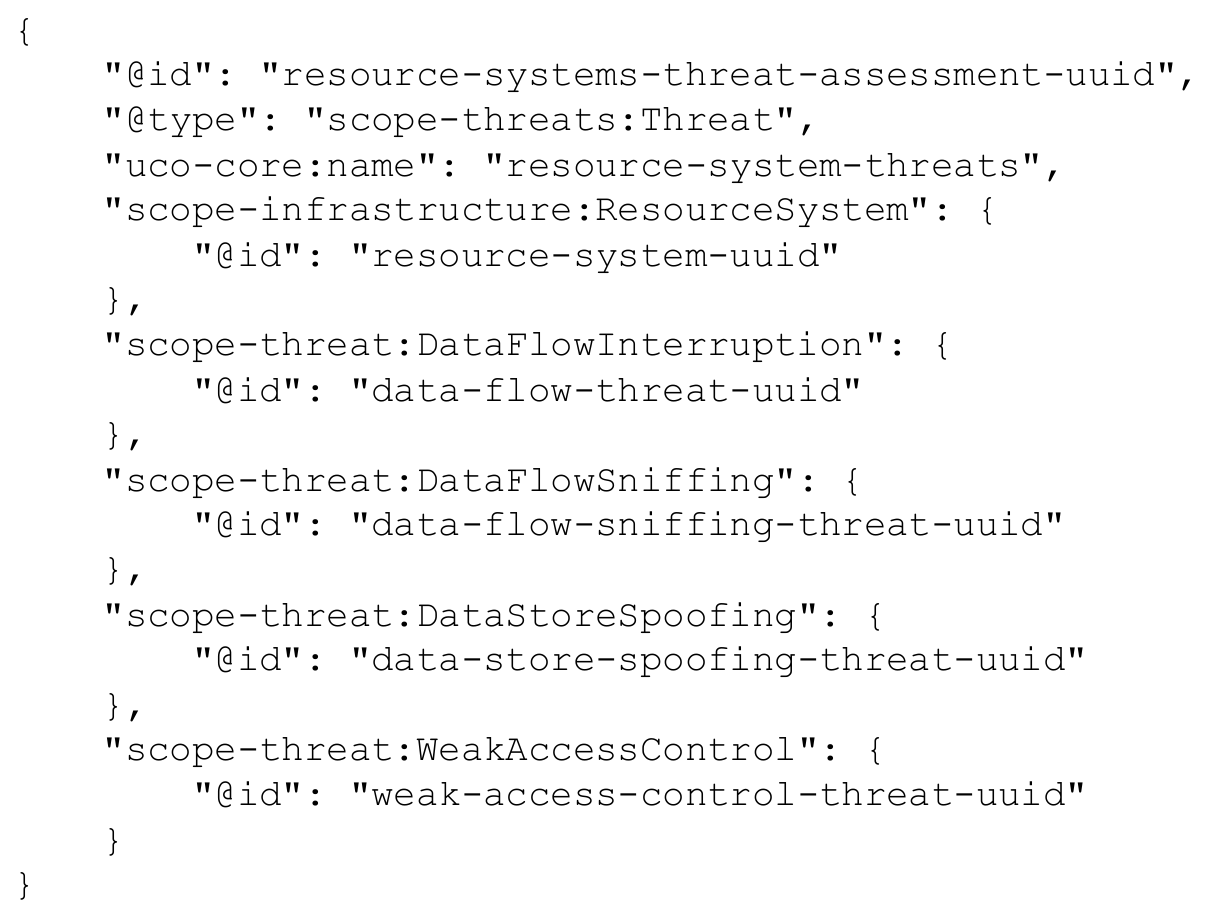}
\centering
\caption{Types of Possible Threats Within Resource System Component of SCI}
\label{fig:SCOPE-SCI-2}
\end{figure}

We can also use \SP to showcase the types of possible threats in SCI. For example, some threats applicable to the Resource System component of a SCI are illustrated in~\autoref{fig:SCOPE-SCI-2}. The component and threats are further assigned identifiers for ease of usage when evidence corresponding to threats and the Resource System component is discovered.

\begin{figure}[htbp]
\includegraphics[width=\columnwidth]{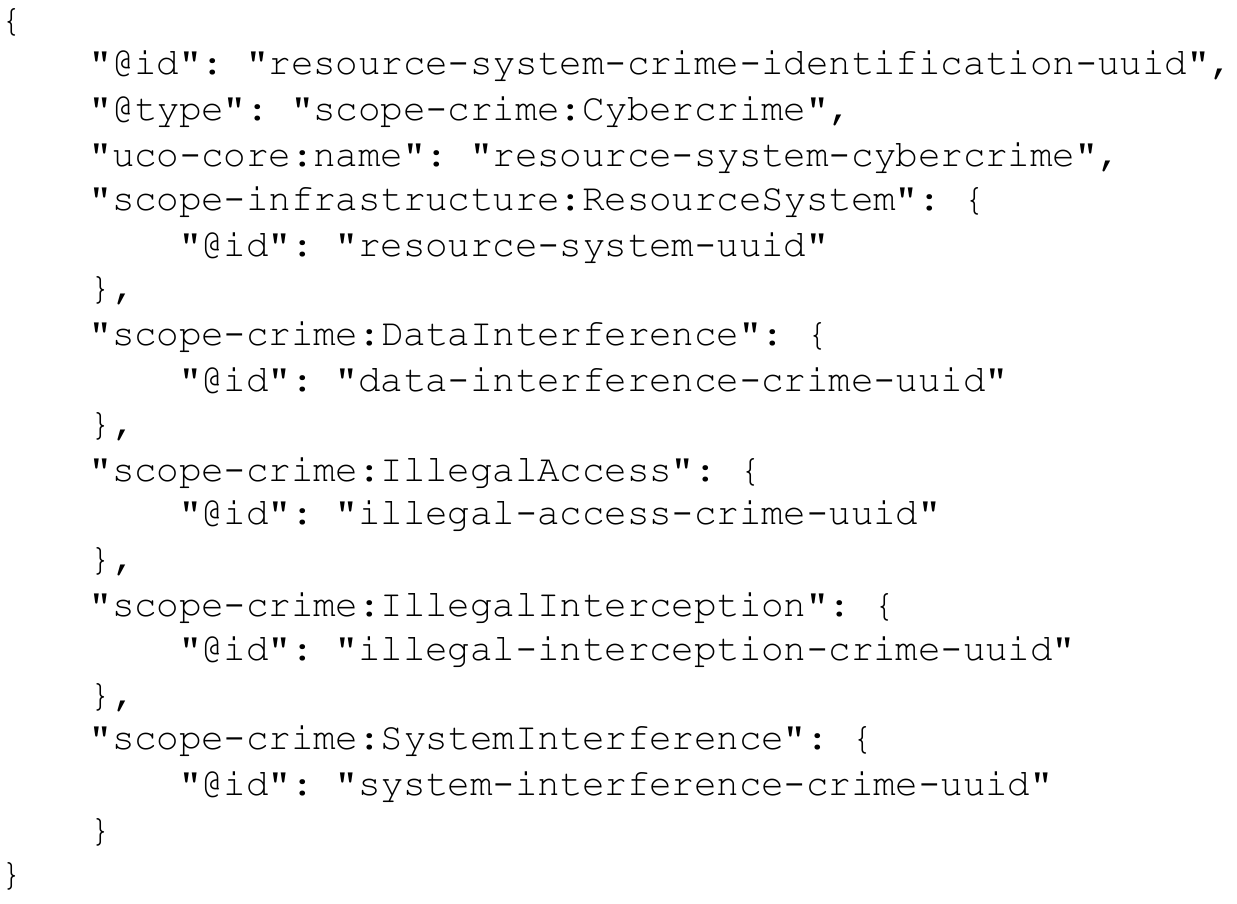}
\centering
\caption{Types of Possible Cybercrime Within Resource System Component of SCI}
\label{fig:SCOPE-SCI-3}
\end{figure}

Using \SP, we then represent the types of cybercrime 
(with reference to~\autoref{fig:SCOPE-SCI-3} and from prior research~\cite{TOK_2023}) that could 
happen within a component of the SCI (Resource System component in this example). In this case, possible cybercrime, such as data and system interference and illegal access and interception, are shown. 

\begin{figure}[htbp]
\includegraphics[width=\columnwidth]{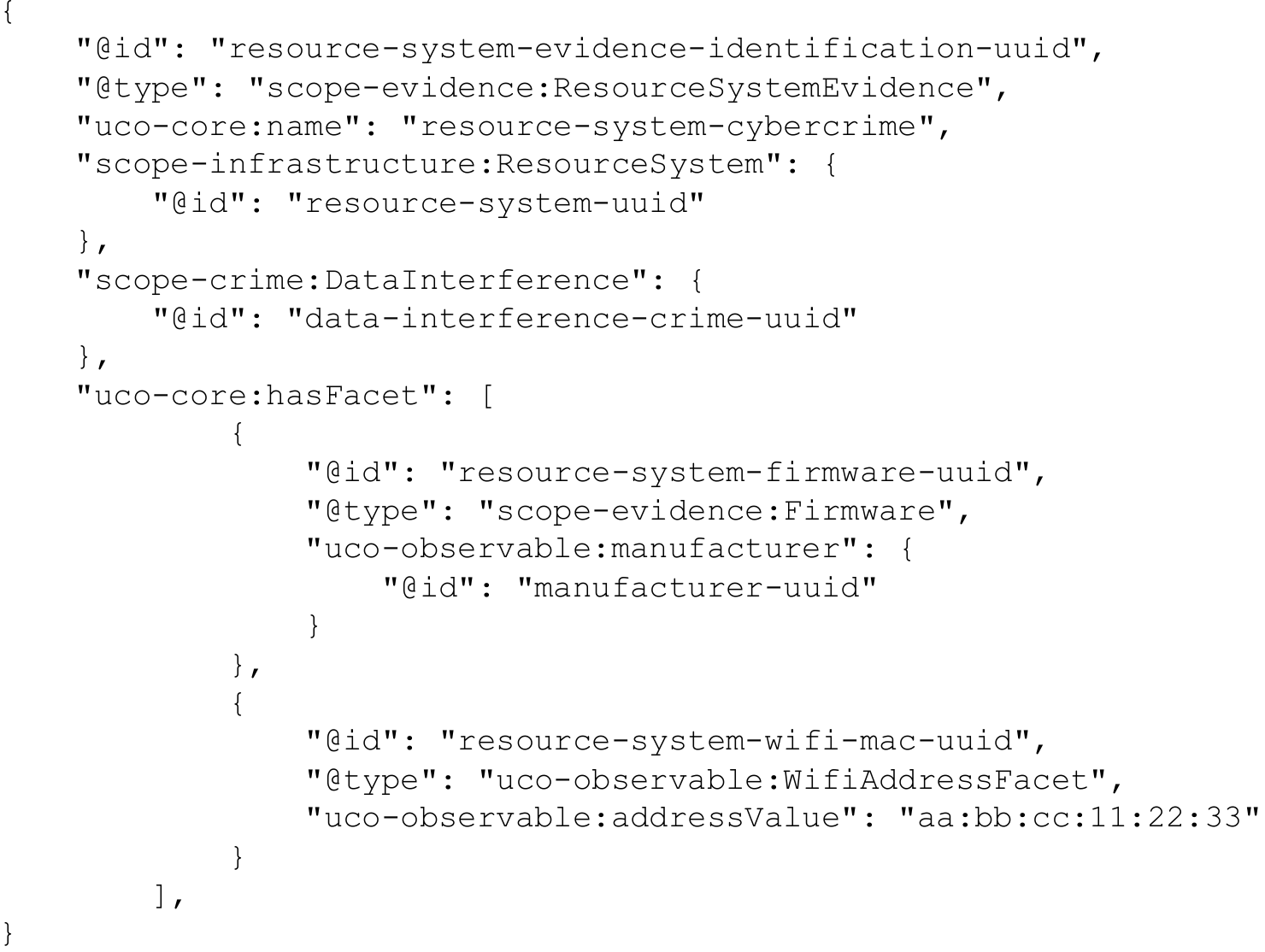}
\centering
\caption{Digital Evidence Mapped to Cybercrime Within Resource System Component of SCI}
\label{fig:SCOPE-SCI-4}
\end{figure}

We could also use \SP to map the evidence captured or needed for the respective cybercrime. For instance, it was discovered that a data interference cybercrime occurred within the Resource System of a SCI. Further investigation indicated that an internet-facing hardware firmware component was compromised, and the component's manufacturer and Media Access Control (MAC) address was recorded. The corresponding details are presented in~\autoref{fig:SCOPE-SCI-4}).

\begin{figure}[htbp]
\includegraphics[width=\columnwidth]{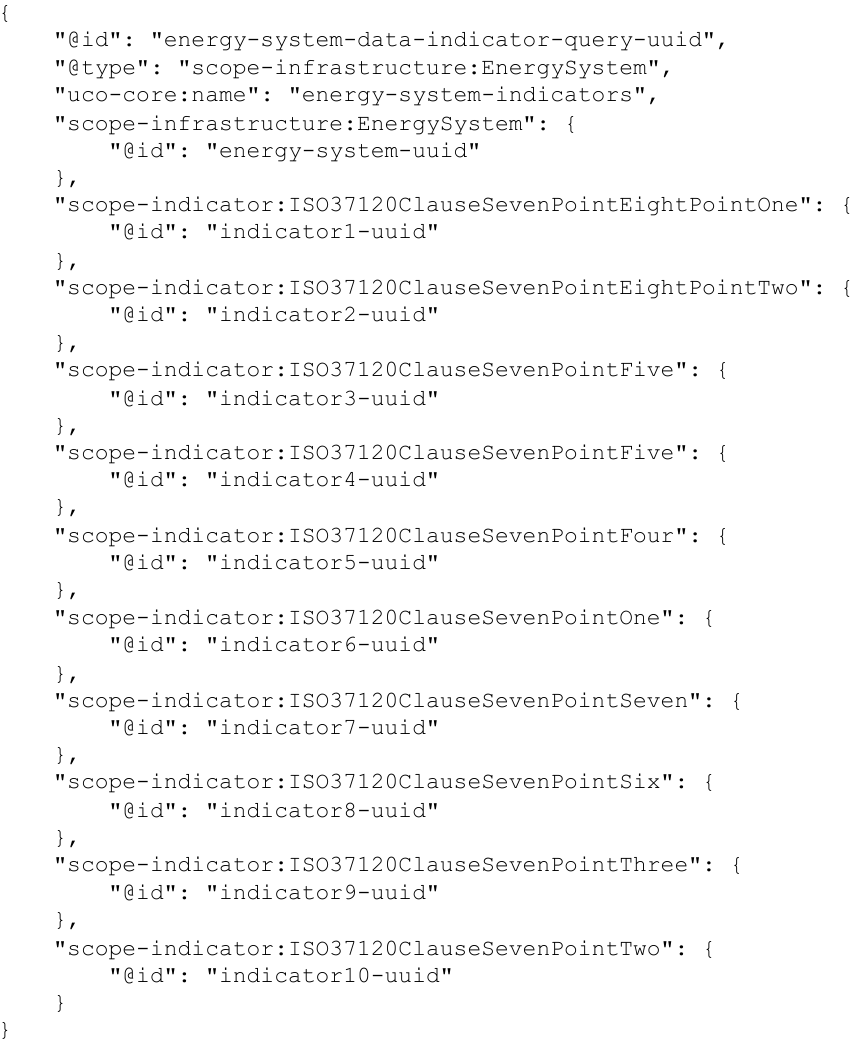}
\centering
\caption{Data Indicators Within Resource System Component (Energy System) of SCI}
\label{fig:SCOPE-SCI-5}
\end{figure}

Finally, with reference to~\autoref{fig:SCOPE-SCI-5}, we can use \SP to collate the data indicators within a specified SCI system. 
All data indicators related to the Energy System (from the Resource System grouping of SCI and referencing ISO37120~\cite{ISO_37120_2018}) are grouped and assigned individual identifiers. 
% \todo{These data indicators are from an ISO I guess, 
% please reference the ISO.} - done (30/07/2024)
This allows a quick enumeration of all associated data indicators of a larger system, which could further assist investigators to ascertain about the affected data outputs %could be affected 
when SCI is subject to an attack or cybercrime.

In summary, the examples using \SP demonstrate the additional SCI-related functionalities added to UCO/CASE. By adapting \SP into the UCO/CASE ontology instances, DFI and LEA can immediately represent SCI-related threats, cybercrime and data components without additional time to construct SCI-related components. In addition, the SCI represented within \SP is technology-agnostic, thus making it highly compatible with diverse SCI systems.
% 4th section

\section{Evaluation of \SP} 
\label{SCOPE_Evaluation}

For the purposes of determining the efficacy of {\SP}, we developed scenarios that emulate real-life investigations. Such scenarios 
would take place during a cybercrime investigation in a smart city, particularly those that address the aims and purposes of the United Nations Sustainable Development Goals and data indicators outlined by Tok and Chattopadhyay~\cite{TOK_2023}. The large overarching scenario is split into three smaller scenarios representing key activities during a cybercrime investigation - \raisebox{.5pt}{\textcircled{\raisebox{-.9pt} {1}}} initial overview of the scenario and its ontology representation, \raisebox{.5pt}{\textcircled{\raisebox{-.9pt} {2}}} incident investigation with Tactics, Techniques and Procedures (TTPs) identified during the examination and \raisebox{.5pt}{\textcircled{\raisebox{-.9pt} {3}}} containment and recovery using identified Indicators of Compromise (IoC). 

\begin{figure}[H]
\includegraphics[width=\columnwidth]{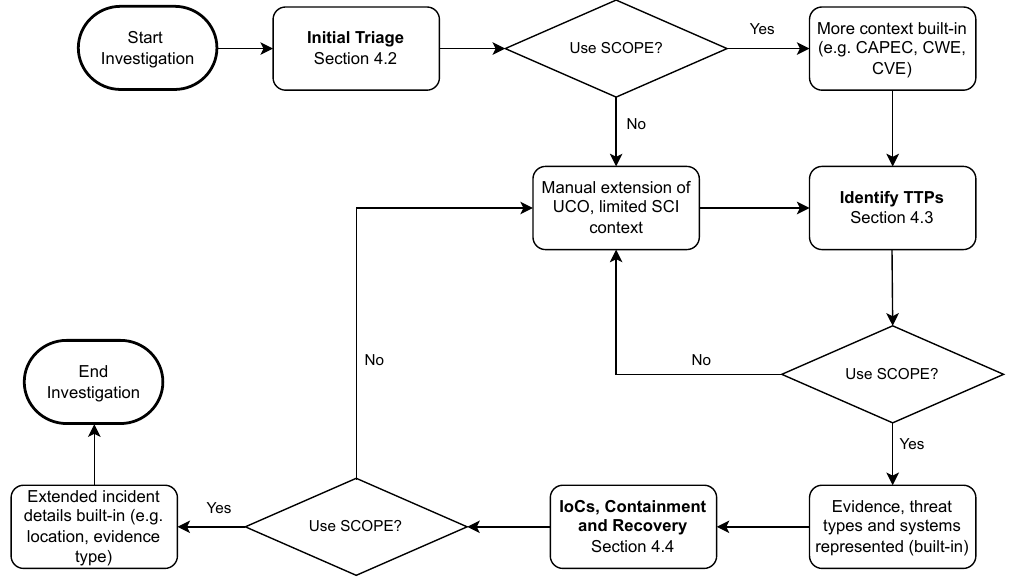}
\centering
\caption{Evaluation Scenario Flowchart}
\label{fig:SCOPE-Scenario-Flowchart}
\end{figure}

These scenarios come with corresponding evidence necessary for the evaluation exercise. The evidence is manually mapped by researchers to the corresponding ontologies being evaluated. In the case of \SP, the researchers also manually mapped threats and evidence to the ATT\&CK techniques and CAPEC attack patterns. To ensure consistency and to validate this mapping, two researchers independently performed the mapping and reached a consensus. The digital evidence in the scenarios is presented in both ontologies (UCO/CASE and \SP) to demonstrate investigation and collaboration workflow optimizations for DFIs. To compare \SP with a baseline, we manually extended the UCO/CASE ontology used in the evaluation scenario to incorporate SCI elements. Such efforts would require DFI involved in the investigation to be proficient in SCI threats, digital evidence types, attack techniques and classifications. Moreover, significant modifications and time are required to prepare this baseline. This reflects the need for a novel extension from UCO/CASE that covers SCI use cases, and DFI can directly adopt it for cybercrime-related incidents in SCI. We provide a visual summary of the evaluation exercise, along with salient points where \SP empowers DFI in~\autoref{fig:SCOPE-Scenario-Flowchart}.

\subsection{\SP Scenario Overview}  
\label{SCOPE_Scenario_Overview}

Our proposed scenario is based on an actual APT Group (APT41, alternatively known as Brass Typhoon or Barium) and draws on several TTPs listed by MITRE~\cite{APT41_MITRE_2024}. The scenario is instantiated geographically in Singapore and in a township (Punggol) that has been announced as a smart district publicly (Punggol Digital District)~\cite{JTC_2022}. APT41 was chosen as the reference adversary for the scenario as it targets high-technology industries, and some of its activities have been attributed to victims in Singapore~\cite{APT41_MANDIANT_2022}.

In the context of the scenario, we have established a few parameters. Firstly, just like the current events, cybersecurity is a fast-paced and constantly evolving field with new technological developments and obsolescence occurring. Homonyms are increasingly common in computer science and cybersecurity, where exact definitions and standardization are essential for collaboration and investigation. Furthermore, tracking the growth of IoT development and the numerous sensors and devices being developed is increasingly difficult.

As such, the second parameter will be that first responders will find it challenging to identify and communicate what has already been discovered. This parameter is also exacerbated by the fact that the new Punggol Digital District had been built to develop and integrate smart solutions, resulting in an oversized amalgamation of interconnected technology systems that are built towards the envisioned Smart City as defined by Tok and Chattopadhyay~\cite{TOK_2023}.

In the final parameter, we presume that an incident investigation from a professional Cybersecurity Incident Response Team (CIRT) adheres to an established framework, such as those from the National Institute of Standards and Technology (NIST)~\cite{NIST_sp.800-61r2_2012, NIST_sp.800-61r3_2024}. It usually follows the same process and procedures, such as evidence acquisition, chain of custody and investigations in an analysis facility while striving to achieve cybersecurity outcomes such as detection, response and recovery. Most CIRTs will maintain similar standard operating procedures with some workflow tweaks depending on their mission, goals and requirements.

\subsubsection{Prologue}
\textit{On 1\textsuperscript{st} January 2100, the Singapore Computer Emergency Response Team (SingCERT) was notified from various organizations and companies situated at the Punggol Digital District that they were struck by a ransomware attack. APT Triple Dragon has claimed responsibility for the attack and has demanded a ransom payment of US\$250 million to be made in 7 days.}

\subsection{Scenario 1 - Initial Triage (1\textsuperscript{st} January 2100)} \label{SCOPE_Scenario_1}

SingCERT dispatched a team of cyber incident responders equipped with jump kits and commenced initial triage of the incident. The incident responders performed the following actions:

\begin{enumerate}
	\item \textbf{Imaging affected devices.} The incident responders used Tableu Forensic Imager TX1 to image the storage devices of infected computers for endpoint-based evidence.
	
	\item \textbf{Initial analysis.} As a complete image of devices could take time, an initial assessment is also required to determine the impact and extent of damage. Using specially configured portable removable devices with triaging tools such as memory capture and assorted forensic artifact acquisition tools, the responders performed initial on-the-spot analysis.   
	
	\item \textbf{Network traffic analysis.} Relying on endpoint logs and device artifacts in a ransomware incident could be insufficient. Network packet capture and network traffic analysis provide a different perspective in investigations as valuable data, such as domains contacted and remote commands, could be captured to gain a deeper insight into the incident.
	
	\item \textbf{Interim assessment.} Based on the previous steps, the incident responders gather, collaborate and corroborate their findings to make an interim assessment of the incident.
\end{enumerate}

The responders discovered that the extent of the infection was far larger than initially assessed, as the ransomware identified as Tiki could propagate via IoT systems. Thus, in an effort to be meticulous, SingCERT has directed the seizure of all systems that may be linked to the attack.

\subsubsection{Lab Investigation}

After seizure and retrieval of the impacted systems, logs collected were fed into a standalone computer and processed via log ingestion tools such as the Elastic, Logstash and Kibana (ELK) stack~\cite{ELK_2024} and Splunk~\cite{Splunk_2024} for analysts to trawl through. A chain of custody was established and maintained for all seized digital evidence, ensuring the evidence was admissible in court. Additional bit-by-bit duplication was performed on the large data storage devices that could not be acquired during the initial triage due to their size. After the investigation efforts, analysts identified and extracted the ransomware from the infected systems, identifying it as Encryptor RaaS ransomware. 

Throughout the attack, the perpetrator employed sophisticated, tailor-made, and elusive tools that successfully bypassed and circumvented the antivirus programs and traditional security measures. The malware instances examined by SingCERT turned out to be any of the following: 1) novel variants previously unidentified in the wild, 2) undetected by the conventional anti-malware systems utilized by the various security operation centers affected, 3) new variants of older malware, 4) custom tools never encountered previously or reported by security vendors or a combination of modified open-source tools designed to conceal the attacker's activities.

The attacker is evaluated as proficient and advanced, displaying traits indicative of an APT group, particularly taking into account the previous characteristics observed from the evidence.

\subsubsection{Impact}

Initial triage by the responders estimated the damage to be merely the infected on-site servers and end-user computers, with key hosts being ransomed. However, lab analysis of network traffic revealed that infection has spread much further than initially presumed. The incident responders identified malicious domains, and even IoT devices and cloud-based systems were revealed as being compromised, 
as they communicated with these malicious domains. 

\subsubsection{Timeline}

The attack began in June 2099, as certain compromised hosts were identified to have made successful callbacks to malicious domains from early June. Host logs have determined that the point of entry was via spear phishing and an unsecured, exposed public-facing development website.

From June to December, the attacker was identified as propagating and spreading through the network and exfiltrating critical information from compromised data servers. Furthermore, it has been discovered that compromised game companies infected their in-development games with malicious backdoors and distributed them via video game digital distribution services and online storefronts to other victims.

\subsubsection{Evaluation of Ontologies Used for Scenario 1}

\begin{figure}[htbp]
\includegraphics[width=\columnwidth]{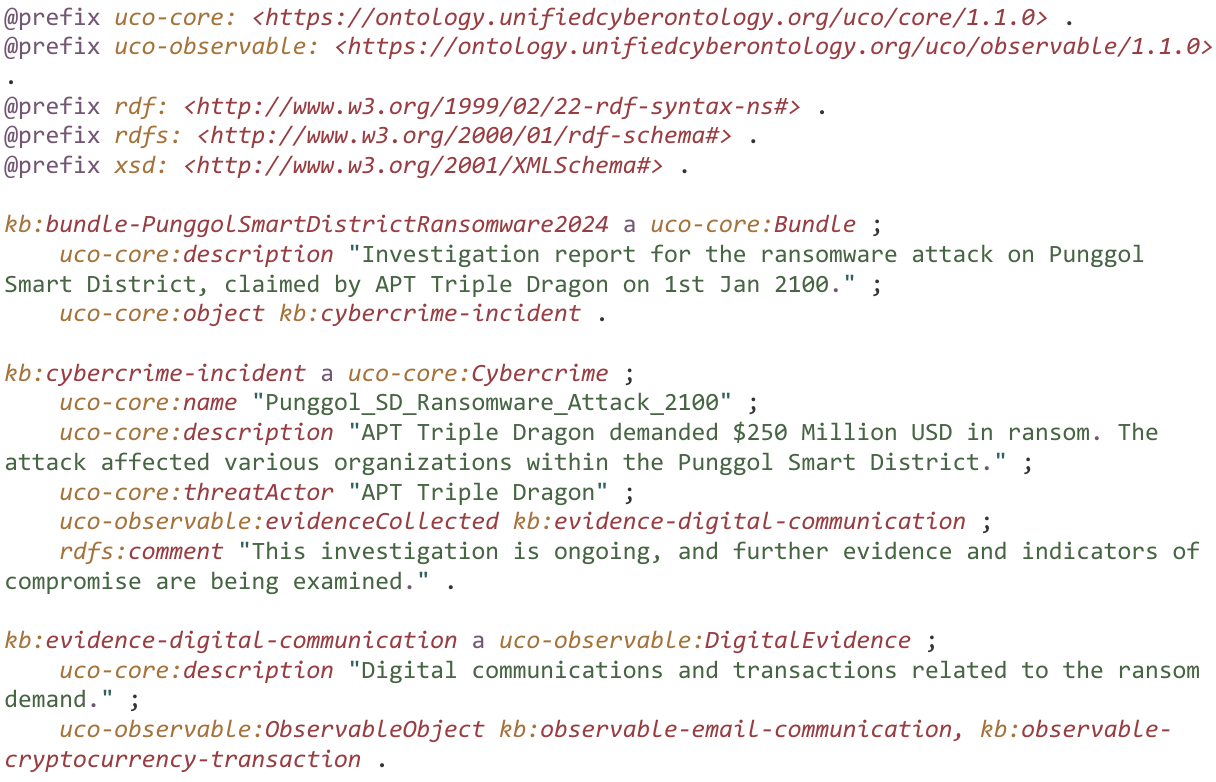}
\centering
\caption{UCO Representation of Incident}
\label{fig:UCO-RDF}
\end{figure}

\begin{figure}[htbp]
\includegraphics[width=\columnwidth]{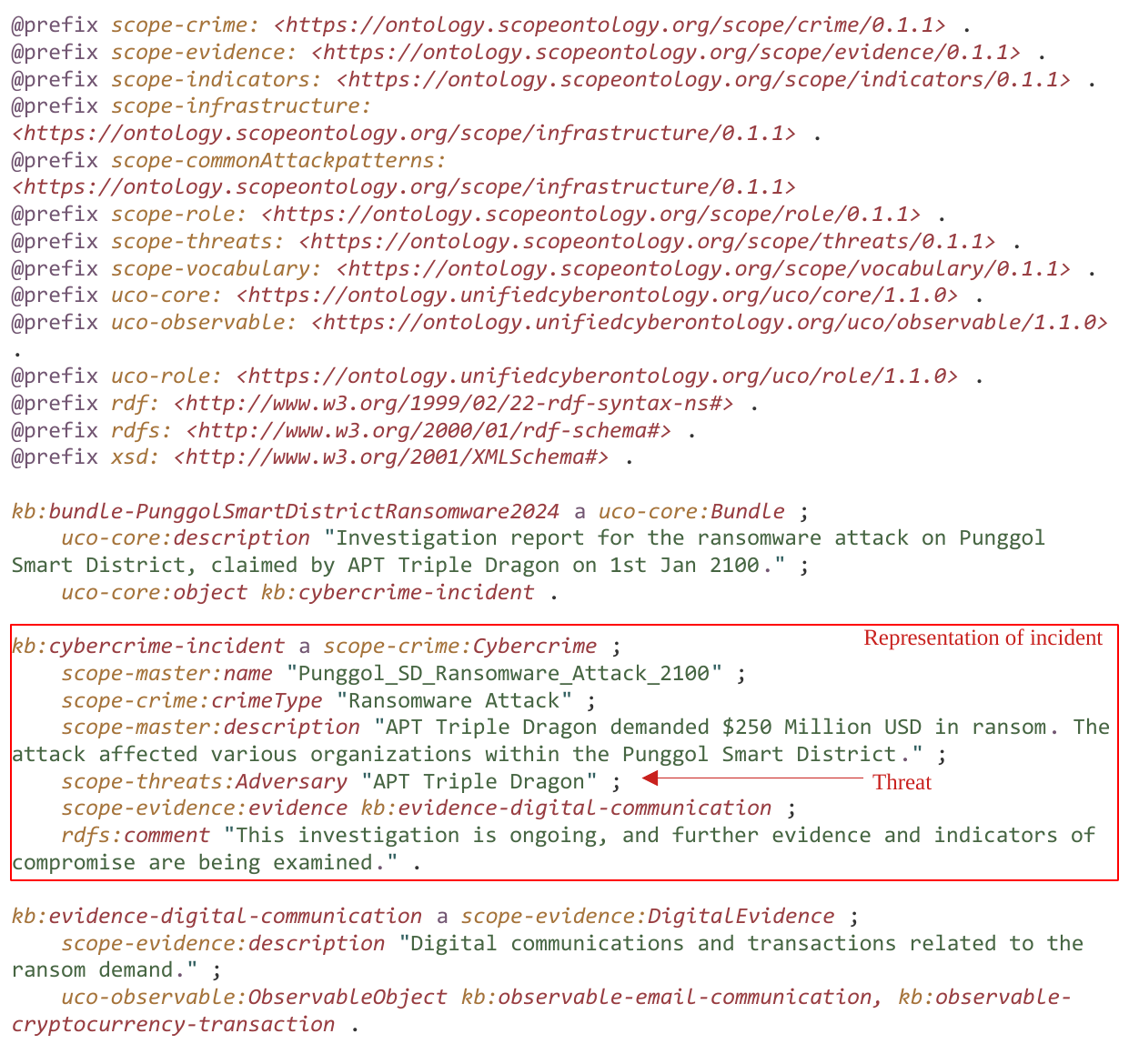}
\centering
\caption{\SP Representation of Incident}
\label{fig:SCOPE-RDF}
\end{figure}

\autoref{fig:UCO-RDF} and~\autoref{fig:SCOPE-RDF} represent the scenarios in resource description framework format using  
UCO \& CASE and  using \SP, respectively (differences are highlighted by the red box as shown in~\autoref{fig:SCOPE-RDF}). 

Both ontologies are  capable of representing Scenario 1. However, the \SP ontology provides further smart city infrastructure-specific descriptions (e.g., \textit{crimeType}, \textit{Adversary}) which allow an increased level of granularity as well as adding in more information fields that could add useful context for cybercrime investigations. A greater distinction between \SP and UCO/CASE will be demonstrated in the subsequent sections and scenarios.

% Start of Scenario 2

\subsection{Scenario 2 - TTPs Identified} \label{SCOPE_Scenario_2}
Based on industry best practices for determining and classifying TTPs used by adversaries (including APTs), the TTPs employed by APT Triple Dragon are categorized and classified according to the relevant MITRE ATT\&CK techniques~\cite{ATTCK_MITRE_2024} and technique identifiers are outlined in the following sub-sections. The information will then be represented via UCO/CASE and \SP.

\subsubsection{Initial Access}  \label{Scenario_2_Initial_Access}
\paragraph*{\textbf{Exploit Public Facing Application (T1190)}} The attackers initially targeted a vulnerable, public-facing gaming development website, leveraging an exposed GitLab console. This vulnerability, identified as CVE-2022-2884, allowed them to execute code remotely, gaining an initial foothold in the network.

\paragraph*{\textbf{Spearphishing Attachment (T1566.002)}} Concurrently, a spearphishing campaign was launched, targeting specific individuals within the organization. Malicious documents sent via email served as the initial dropper for malware, deceiving users into enabling macros or executing the embedded code, thus bypassing traditional email filters and security measures.

\subsubsection{Discovery} \label{Scenario_2_Discovery}
\paragraph*{\textbf{File and Directory Discovery (T1083)}} After gaining initial access (with reference to Section~\ref{Scenario_2_Initial_Access}), the attackers conducted reconnaissance to identify valuable files, directories, and configuration settings as seen in host logs. This allowed the attackers to identify the layout of the network, identifying targets for lateral movement, and locating sensitive data for exfiltration.

\paragraph*{\textbf{System Information Discovery (T1082)}} Gathering information about the operating systems, software installations, and network configurations would be crucial for tailoring further attacks, exploiting specific vulnerabilities, and avoiding detection.

\subsubsection{Execution}
\paragraph*{\textbf{Command and Scripting Interpreter (T1059)}} Leveraging on information gleaned from initial access and discovery (referencing Section~\ref{Scenario_2_Initial_Access} and \ref{Scenario_2_Discovery}), the attackers utilized command-line interfaces and scripting to execute their payloads. This included the deployment of ransomware and Remote Access Trojans (RATs), allowing them to maintain control over compromised systems and perform further malicious activities undetected.

\subsubsection{Persistence and Privilege Escalation}
\paragraph*{\textbf{External Remote Services (T1133)}} By exploiting the exposed GitLab console (from Section~\ref{Scenario_2_Initial_Access}), attackers ensured persistent access to the compromised web server. This access facilitated the lateral movement across the network and the compromise of additional hosts.

\paragraph*{\textbf{Valid Accounts (T1078)}} Through password spraying attacks, the attackers gained access to valid user accounts, escalating their privileges within the network. This allowed them to access critical resources and data, further entrenching their presence.

\subsubsection{Lateral Movement}
\paragraph*{\textbf{Remote Services: SSH, RDP (T1021)}} With valid accounts at their disposal, attackers used services like SSH (Secure Shell) and RDP (Remote Desktop Protocol) to move laterally across the network, accessing and compromising additional systems.

\paragraph*{\textbf{Lateral Tool Transfer (T1570)}} To facilitate their movement and maintain access within the network, the attackers transferred tools or malware from one compromised host to another. This technique allowed for the execution of specific payloads tailored to each target system.

\subsubsection{Defense Evasion}
\paragraph*{\textbf{Obfuscated Files or Information (T1027)}} The malware deployed during the attack was meticulously designed to evade detection by endpoint security solutions. Through obfuscation and packing techniques, including the use of virtual machine detection evasion tactics, the attackers minimized the malware's footprint and avoided triggering security alerts.

\subsubsection{Collection }
\paragraph*{\textbf{Automated Collection (T1119)}} Malicious scripts and custom malware that automatically collects specified types of documents and data from the compromised systems were identified to have harvested important financial information.

\paragraph*{\textbf{Input Capture: Keylogging (T1056)}} Deploying keyloggers enabled the capture of credentials, sensitive information, and other inputs directly from the users' keystrokes, further compromising personal and organizational data.

\subsubsection{Command and Control}
\paragraph*{\textbf{Application Layer Protocol: Web Protocols (T1071)}} The attackers utilized HTTP, HTTPS, or other web protocols for command and control (C2) communications which blended the malicious traffic with legitimate web traffic, and made detection more challenging.

\paragraph*{\textbf{Dynamic Resolution (T1568 \& T1071.004)}} The attackers employed techniques such as Domain Generation Algorithms (DGA) and fast flux, making their C2 infrastructure more resilient and challenging to disrupt. Furthermore, they used the Domain Name System (DNS) application layer protocol to avoid detection filtering by blending in with existing traffic with expired domains that they had bought and could already pass through the network firewall. Commands to the remote system and the results of those commands were identified to be embedded within the protocol traffic between the client and server.

\subsubsection{Exfiltration}
\paragraph*{\textbf{Exfiltration Over C2 Channel (T1041)}} Leveraging the established C2 channels to exfiltrate stolen data silently ensures that attackers maintain a low profile while continuously siphoning off sensitive information.

\paragraph*{\textbf{Scheduled Transfer (T1029)}} The attackers had set up automated exfiltration processes that operate at scheduled times can help avoid detection by blending in with standard network traffic patterns, especially during peak hours.

\subsubsection{Impact}
\paragraph*{\textbf{Data Encrypted for Impact (T1486)}} The deployment of ransomware to encrypt critical systems and data disrupts operations and serves as a direct method for financial gain through ransom demands from the APT.

\paragraph*{\textbf{Endpoint Denial of Service (T1499)}} The identified attacks on individual high-value endpoints, such as resource hijacking and request floods, rendered devices unusable, further complicating recovery efforts.

\subsubsection{Evaluation of TTPs Representation (Scenario 2)}
\begin{figure}[htbp]
\includegraphics[width=\columnwidth]{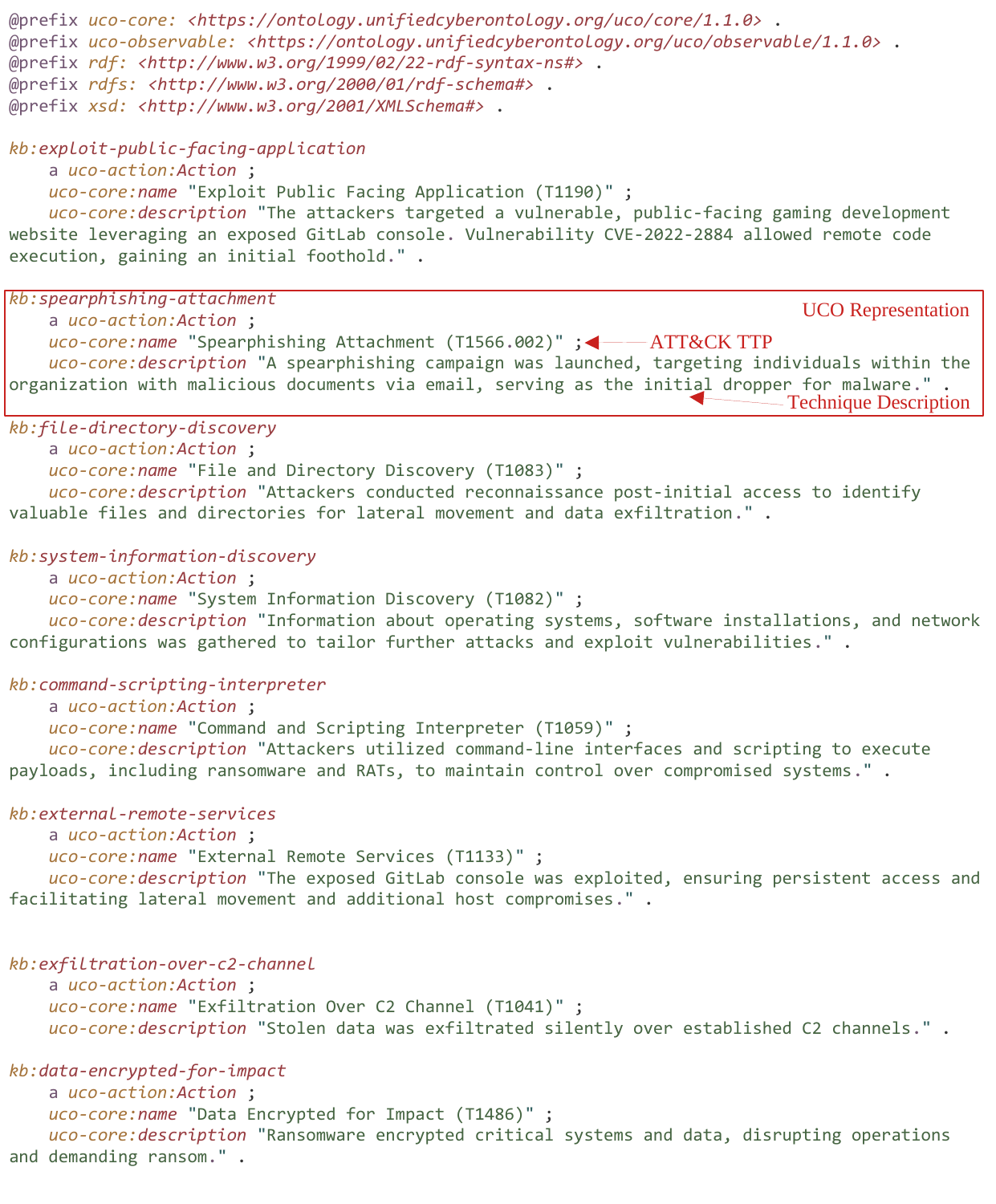}
\centering
\caption{UCO Representation of Tactics, Techniques, and Procedures}
\label{fig:SCOPE-SCI-6}
\end{figure}

\begin{figure}[htbp]
\includegraphics[width=\columnwidth]{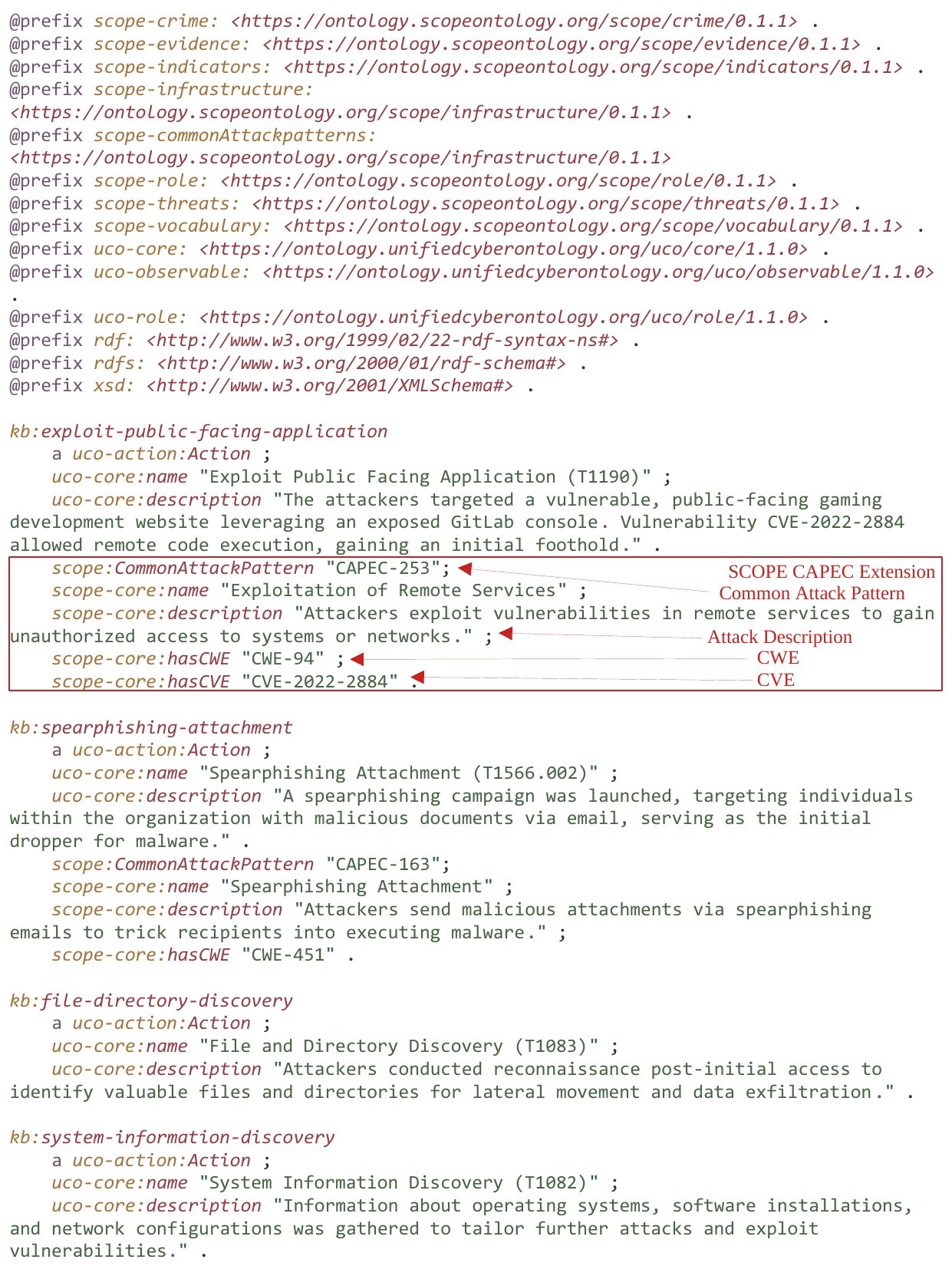}
\centering
\caption{\SP Representation of Tactics, Techniques, and Procedures}
\label{fig:SCOPE-SCI-7}
\end{figure}

Represented in~\autoref{fig:SCOPE-SCI-6} (UCO representation of tactics, techniques, and procedures) and~\autoref{fig:SCOPE-SCI-7} (\SP representation of tactics, techniques, and procedures) were the two different resource description framework representations of the various tactics, techniques, and procedures identified to be used by the threat actor in the scenario. Both have their use cases and help convey essential knowledge about what the attacker has performed to other investigators. In the UCO representation, we demonstrated extending it using the MITRE ATT\&CK framework (knowledge of the ATT\&CK framework would be a prerequisite for DFI involved in the investigation). Meanwhile, in \SP, we further imbued the ontology with MITRE Common Attack Pattern Enumeration and Classification (CAPEC)~\cite{CAPEC_MITRE_2024}.

The different approaches were due to the use case of the \SP ontology. MITRE ATT\&CK provides a structured and comprehensive framework for understanding TTPs that adversaries may use; it focuses on specific actions that may happen at each stage of an attack lifecycle and is more detailed and commonly preferred by security professionals such as threat intelligence analysts. However, with the addition of CAPEC, DFI can enumerate and classify common attack patterns used by adversaries. This allows for a higher level of understanding and documentation of attacks, enabling more effective use in threat modelling and informing of defensive strategies for decision makers. 

Throughout the entire attack, the attacker performs a variety of actions. These actions are reflected and captured using MITRE ATT\&CK as seen in ~\autoref{fig:SCOPE-SCI-6} using UCO, which demonstrates and highlights the cyber adversaries' tactics and techniques. \SP, when adopted on top of UCO/CASE, will also highlight the common attack patterns the defenders observe. \SP also highlights the CWE and CVE as observed in ~\autoref{fig:SCOPE-SCI-7}, noting the weaknesses that the attacker used. This allows the remediation team to take effective action and patch the specific vulnerabilities exploited.

\subsection{Scenario 3 - Indicators of Compromise, Containment and Recovery} \label{SCOPE_Scenario_3}

IoCs that may prove helpful for defenders to denylist or further investigate are expected to be discovered during the investigation. During this fictional scenario, SingCERT discovers that the APT enlists the use of Encryptor Malware, which makes callbacks to various malicious domains. Table~\ref{5_table:3} shows the associated IoCs obtained after malware reverse-engineering was performed on the retrieved samples. 

\begin{table} [htbp]
\caption{Malware IoCs for Scenario 3}
\centering
\begin{tabu} to 0.9 \columnwidth {|m{1.5cm}|m{3.5cm}|m{2.4cm}|}  
    \cline{1-3}
    % \multicolumn{1}{|c|}
    \vspace*{6pt}\textbf{Type of Malware} \vspace*{6pt} & \vspace*{6pt} \hfil\textbf{MD5 Hash} \vspace*{6pt}& \hfil\textbf{Malicious Domains}   \\
    \hline
    \vspace*{6pt}Encryptor Malware\vspace*{6pt} & \vspace*{6pt}00c4c3946ec03c915cfe4cbddff
    e93da\newline f84d54b351b7926106ef377b06
    423734\newline 762a96d79e747457e086e68128
    16b0aa\newline \vspace*{6pt} & \vspace*{6pt}5jua3omslrbkks4c[.]
    onion[.]link\newline agegamepay[.]com\newline ageofwuxia[.]com\newline ageofwuxia[.]info\newline ageofwuxia[.]net\newline ageofwuxia[.]org\newline  \\ \cline{1-3}
    \end{tabu}
    \label{5_table:3}
\end{table}

\begin{figure}[htbp]
\includegraphics[width=\columnwidth]{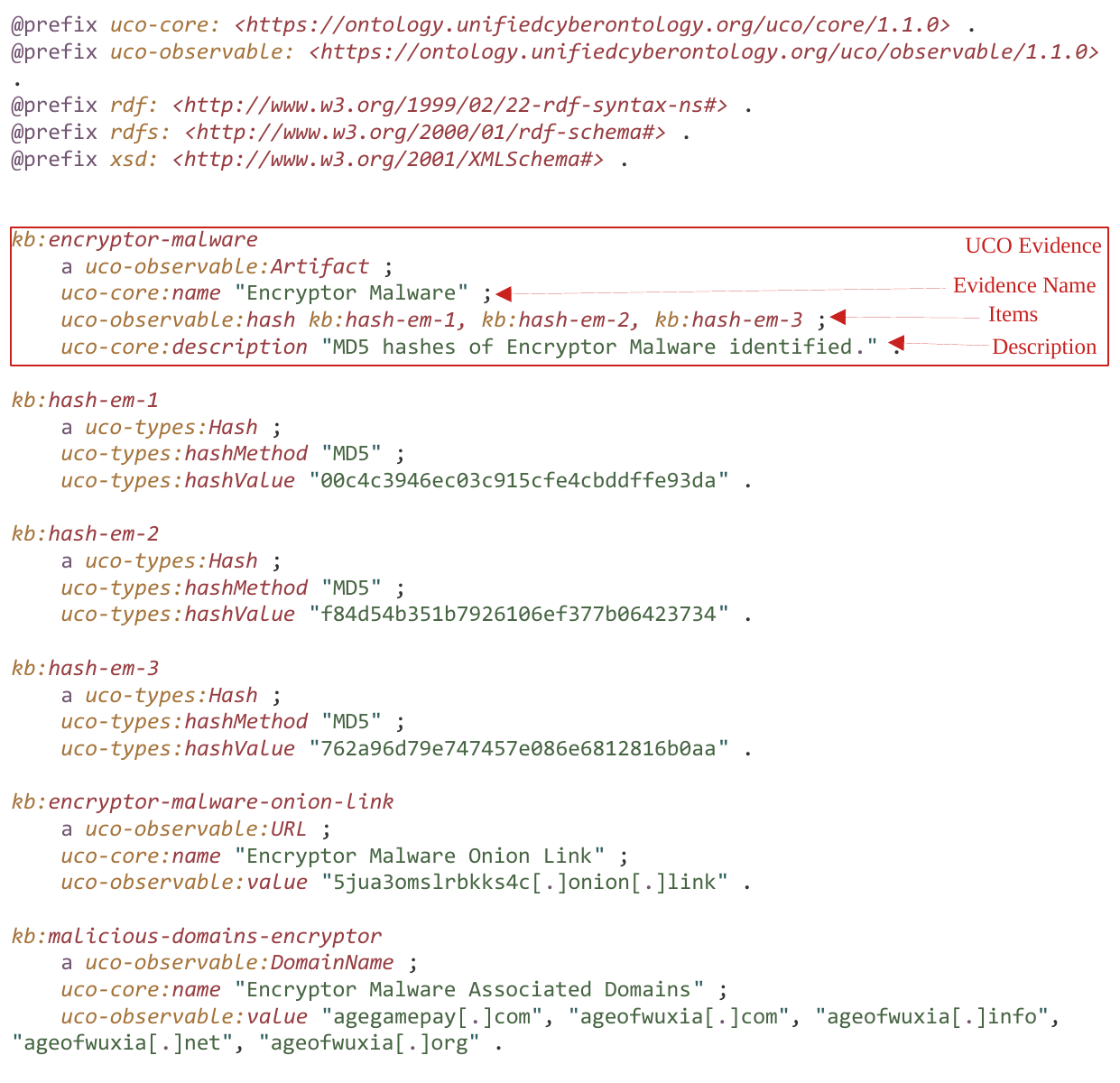}
\centering
\caption{UCO Representation of Evidence}
\label{fig:SCOPE-SCI-8}
\end{figure}

\begin{figure}[htbp]
\includegraphics[width=\columnwidth]{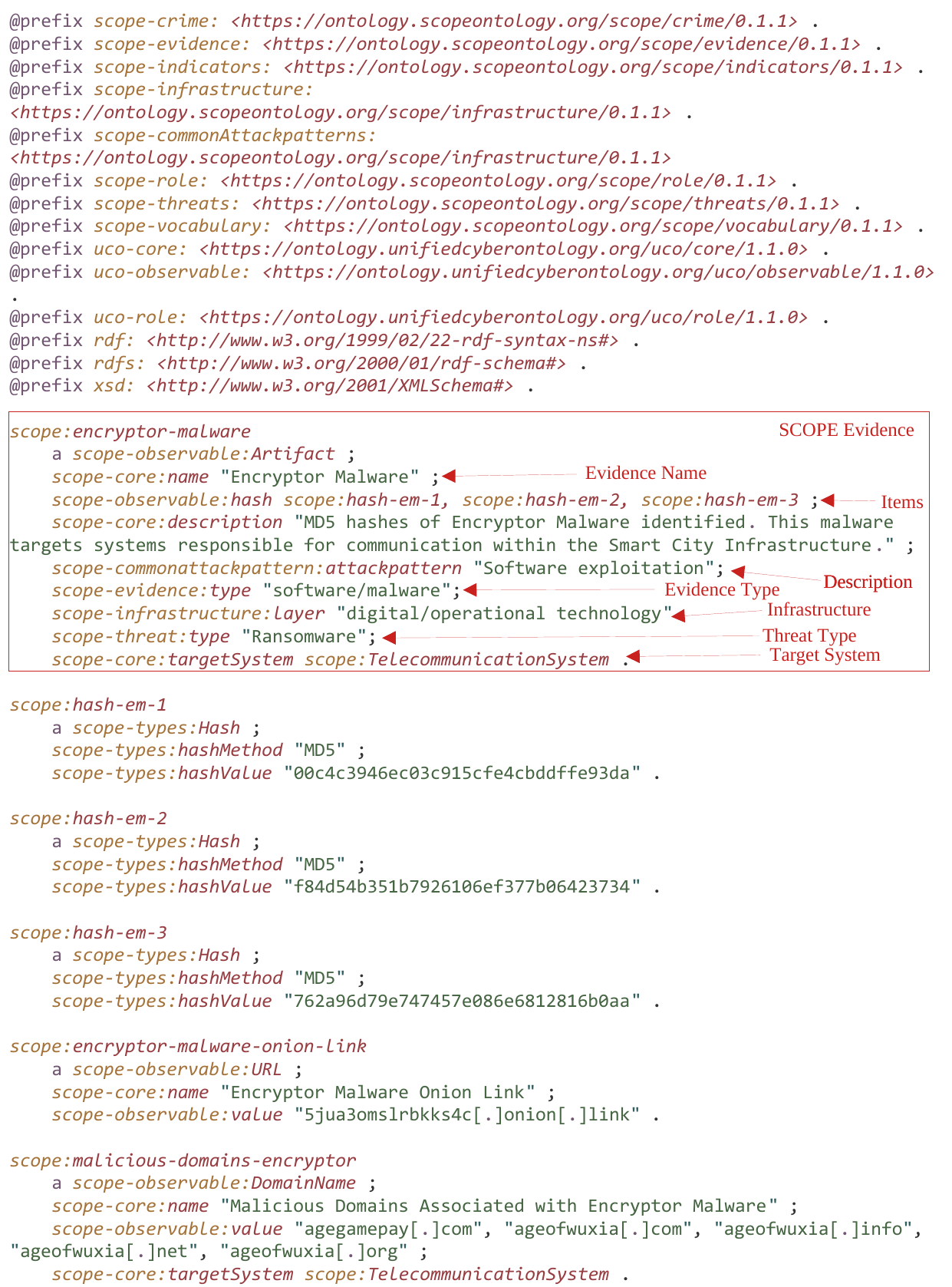}
\centering
\caption{\SP Representation of Evidence}
\label{fig:SCOPE-SCI-9}
\end{figure}

Meanwhile, the data in Table~\ref{5_table:3} were represented using UCO in~\autoref{fig:SCOPE-SCI-8} and via \SP in~\autoref{fig:SCOPE-SCI-9}.

Throughout the investigation process, the IoCs gathered would be helpful for attribution and denial lists used by defenders.~\autoref{fig:SCOPE-SCI-8} displays what the observed indicators are, using UCO/CASE, which is helpful in clearly stating what their origins are and what type of indicators they may be. \SP expands on this information by adding the infrastructure (e.g. \textit{Digital/Operational Technology Layer}) that the indicator may affect and noting what specific types of threat they are and which systems are affected (e.g. \textit{TelecommunicationSystem}) as seen in ~\autoref{fig:SCOPE-SCI-9}. These details add an additional layer of granularity, which would speed up the investigation process and ensure that critical pieces of information are not overlooked when shared with other analysts during the investigation process.

% \subsection{Scenario 4 - Recovery Post Incident}

As the ransomware ravaged the network and encrypted essential data on the hosts, the SingCERT team immediately mitigated the damage. The infected hosts were swiftly quarantined, and the ransomware was removed to prevent further spread and damage. A patch was quickly developed and deployed with a custom script that detected and eradicated the ransomware to avoid reinfection.

To reduce further damage to the companies and organizations, restoration from backups was performed to restore business activity and ensure that affected victims could continue performing critical functions despite all that had happened. 

While analyzing logs and evidence files from the event, the digital forensic team discovered a symmetric encryption key in transit around four days after the ransomware deployment. The key was found in a network packet capture recorded during the first day of the initial compromise when the attacker started deploying ransomware onto the hosts. 

Due to fortunate circumstances, the infected hosts were all encrypted by the same strain of ransomware and thus used the same encryption key. The SingCERT team used the encryption key, decrypted all the encrypted files, and found that all the data was still intact. This allowed the response team to resolve the ransomware issue without paying a hefty sum to the attackers. The total time taken for recovery took only one day for patches to be deployed, and restoring backups to allow for business continuity, analysis, and encryption key discovery took another three days.

\subsubsection{Evaluation of IoCs, Containment and Recovery Representation (Scenario 3)}

\begin{figure}[htbp]
\includegraphics[width=\columnwidth]{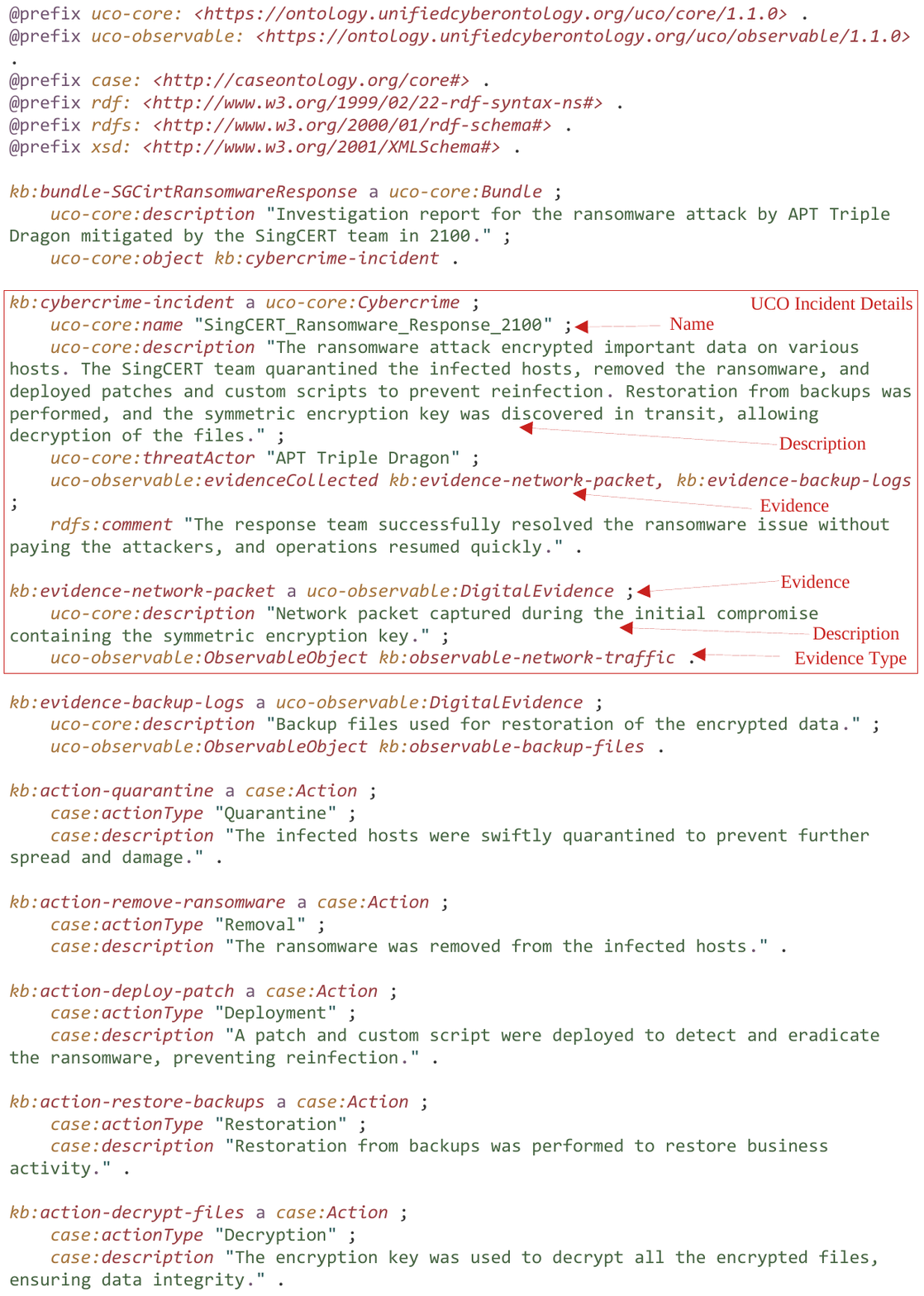}
\centering
\caption{UCO Representation of Recovery}
\label{fig:UCO-RDF-Recovery}
\end{figure}

\begin{figure}[htbp]
\includegraphics[width=\columnwidth]{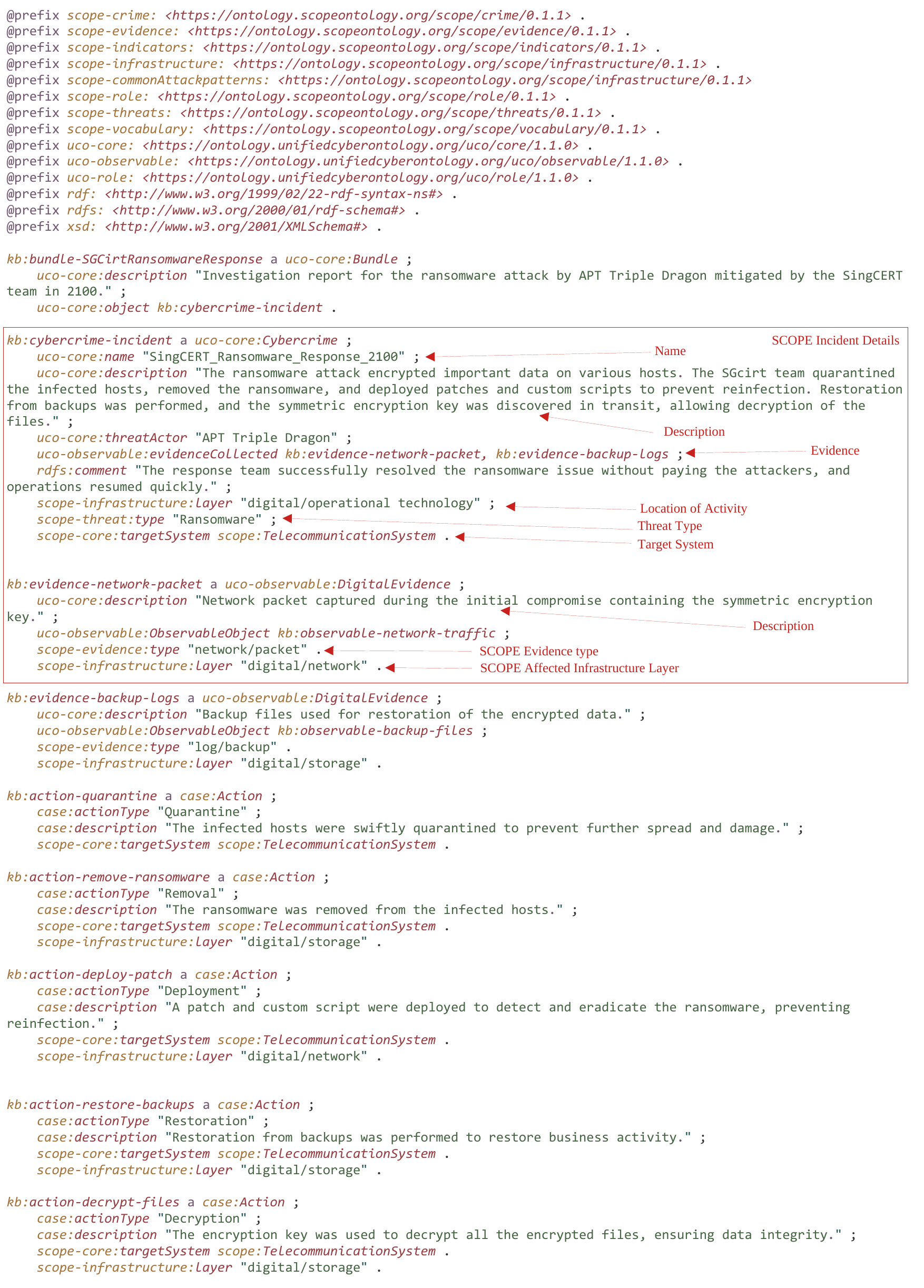}
\centering
\caption{SCOPE Representation of Recovery}
\label{fig:SCOPE-RDF-Recovery}
\end{figure}

The swift recovery process and fortunate discovery of the ransomware encryption key facilitated the restoration of assets affected by the cybersecurity incident. Rapid deployment of patches allowed operations and businesses to resume activities swiftly. The recovery operations are represented by UCO in~\autoref{fig:UCO-RDF-Recovery}, which highlights the actions taken and their descriptions. \SP is represented in~\autoref{fig:SCOPE-RDF-Recovery}, which provides additional context for analysts, such as the location of the activity taken and the evidence type collected. The analysts are empowered to differentiate from SCI compared to other types of environments and systems they may encounter. (Note the red boxes detailing additional contextual information.)
% \subsection{Evidence Representation Evaluation}

With the advent of SCI, it has become essential to distinguish the various attack surfaces, from standard telecommunication systems to transportation systems or operational technology water and energy systems. With \SP, we can represent and share data from emerging technologies, especially complex systems such as smart city developments.

\subsection{Scenario Evaluation Summary}
Through this scenario (with reference to Section~\ref{SCOPE_Scenario_1} to Section~\ref{SCOPE_Scenario_3}), we have identified the key use cases that \SP would value add to cybercrime investigators, allowing users to add more granularity 
% \todo{provide some examples from the scenarios}
in details. An example is seen during the recovery phase of the scenario represented in~\autoref{fig:SCOPE-RDF-Recovery} as compared to~\autoref{fig:UCO-RDF-Recovery}, which highlights the affected areas of infection and damage, allowing increased efficiency and rapid remediation. Complex technical details become easily accessible to other teams who may have a higher level requirement of the ontology, such as malware analysts who may require additional context during the investigation. They may appreciate the extra information such as the affected system and threat type as seen in~\autoref{fig:SCOPE-SCI-9} compared to~\autoref{fig:SCOPE-SCI-8}, which during routine investigations may only capture the basic details of name, hash and filetype. \SP can support evidence and case data sharing through its interoperability with UCO/CASE.

\SP expands on SCI-focused cybercrime, providing specific terminology based on ISO backed definitions which ensures that newer innovations will be covered while retaining older technologies which may still be in use as not all cities in the world develop at the same pace.

% 5th Section

\section{Limitations} 
\label{Limitations}

This section states the limitations of \SP and how we ensured reasonable contributions and innovation to the field. 

\subsection{Industry Adoption}
Cybersecurity developments in academia are generally created independently of industry trends,  with the industry preferring research that may generate revenue compared to purely academic research that may not yield results or are not profitable. Additionally, there may be parallel research being done with industry preferring their own standards over other proposed standardizations, such as using STIX (Structured Threat Information Expression) and TAXII (Trusted Automated Exchange of Intelligence Information) over UCO/CASE, which performed similarly. These frameworks and protocols were developed for cybersecurity information sharing and providing machine readability for automation. However, the industry has indicated a preference towards developments led by leading cybersecurity organizations and companies. 

Despite the lack of adoption of ontologies in the cybersecurity industry, MITRE has also recently developed its own set of cybersecurity-themed ontologies~\cite{kaloroumakis2024d3fend} as more and more organizations have recognized the importance of accurately representing and sharing information. Another notable gap is that \SP aims to provide coverage over SCI, where the technology is still in developmental stages. Thus, there may be no notable adoption or usage in the industry any time soon.

\subsection{Scoping}
Due to the specific pain point we are trying to solve, SCOPE only covers cybercrime in SCI. It thus may lack particular definitions found in other scenarios, such as physical security and social engineering. While this allows us to drill down and ensure coverage of pertinent information, it also leaves the different specialties of cybersecurity unaddressed. Additionally, as SCI is a new and upcoming field, we acknowledge that further developments will occur and render previous efforts useless. Thus, we have tried our best to ensure that \SP is technology-agnostic, allowing it to continue providing coverage as technology develops. Moreover, we have provided 
\SP as an open framework for further development.

\subsection{Threats to Validity}
As an ontology, a possible gap that \SP faces may be content validity. \SP may not contain all relevant concepts and relationships since SCI is a continuously developing field with constant progress in cybersecurity. We can mitigate these issues by conducting extensive and in-depth literature reviews, engaging with experts in the domain to ensure proper coverage as well as continuous and iterative updating of \SP. 
%This will ensure that it will remain up-to-date and relevant. 

Additionally, to ensure that \SP would strike a delicate balance of being granular enough to represent critical points in SCI and broad enough to cover possible scenarios, we have turned to scenario-based testing to ensure that \SP would work in a real-life example.

Another caveat on external and construct validity factors would be that \SP was designed for smart cities in urban environments with advanced technological infrastructure. This may not be representative of rural cities or others that may have a different socioeconomic status. Currently, there are relatively few cities that may be termed smart cities or even cities with the infrastructure that supports developments to become smart cities.

\subsection{Future Developments and Work}
For the future development of \SP, some fields were not utilized that we desire to address in future iterations. Since this was the first version of \SP, fields such as \textit{owl:priorVersion} and \textit{owl:backwardCompatibleWith} were not used, and we have yet to develop a corresponding Python API like CASE. %However, if UCO/CASE is willing to explore \SP, 
Nonetheless, our work is compatible with UCO/CASE projects, 
as we adhered to the corresponding ontology requirements.

Further future developments include aiming to extend Instance-Level Relationship Representation. While ontologies offer robust frameworks for representing structured knowledge in complex domains, they however inherently limit direct representation of relationships between individual instances, such as linking distinct cyber incidents. This gap can be addressed by integrating ontology rule languages like Semantic Web Rule Language (SWRL) or Datalog, which allow advanced reasoning to define and infer relationships at the instance level within the ontology.

We also intend to incorporate SWRL rules or Datalog-based reasoning within \SP to enhance its capability for capturing inter-instance relationships. Such an extension would enable us to model complex patterns, identify recurring behaviors, and establish incident linkages that are essential for forensic analysis. This would advance \SP's utility as a dynamic tool for investigating and analyzing cyber incidents within interconnected systems, including Smart City Infrastructure. Incorporating these rule languages is a significant step toward realizing a fully interconnected knowledge base that supports in-depth, context-aware incident analysis in rapidly evolving cyber-physical environments.

Finally, we plan to conduct a user study with digital forensic professionals from the public and private sectors with respect to the usage of \SP in their investigation processes related to SCI to determine how we can further improve \SP for industry usage. We will also explore creating open-source tools that focus on using \SP as a mechanism for investigation and data sharing.
% 6th section
\section{Related Work} \label{RelatedWork}

Digital forensics investigation has seen significant advances in the methodologies, tools, and techniques used to extract features and correlate evidence. These tools and techniques have various degrees of adoption and perform their objectives to different levels of success. Additionally, to improve collaboration between organizations, efforts have been made to propose a variety of formats that can help standardize the information shared.

\subsection{Feature Extraction and Correlation}
One notable example of innovative developments in feature extraction would be Garfinkel's cross-drive analysis~\cite{GARFINKEL200671}, where he introduces the technique of examining multiple hard drives simultaneously to capture patterns and identify connections that may not be visible in isolation. This approach is beneficial in extensive complex investigations that may span multiple systems.  Flaglien et al.'s~\cite{flaglien2011malware} cross-evidence correlation research on malware also provides a comprehensive framework for malware detection by leveraging cross-evidence correlation to identify malware traces across different hosts. Their proposition increases the speed and accuracy of malware identification and allows efficient processing of large datasets, which may bog down investigators. Another relevant research would be the FACE framework by Case et al.~\cite{CASE2008S65}, which, in addition to automating the process of discovering and correlating digital evidence from multiple data sources, also provides a suite of tools and techniques to analyze information from a myriad of sources and standardizes the extracted data as eXtensible Markup Language (XML) and Attribute Relation File Format (ARFF).

Despite these advances in feature extraction and correlation in digital forensics, a few gaps still need to be addressed. One major limitation would be the need for standardized methods of ingesting data from more diverse sources, such as s Garfinkel's~\cite{GARFINKEL200671} cross-drive analysis, which has robust support but lacks the capability of ingesting network and application logs. Similarly, while efficient at identifying malware, Flaglien et al.'s~\cite{flaglien2011malware} research does not support the holistic depth needed in digital forensics. Another issue would be the need for more real-time analysis in the FACE framework~\cite{CASE2008S65}, paramount in a time-sensitive incident response scenario. These gaps signify the importance of digital forensic frameworks that can integrate well into the existing ecosystem of tools with industry-standard formats such as RDF and the significance of analyzing various sources in real time.

One recent development by  Kougioumtzidou et al.~\cite{Kougioumtzidou_2024} used a neural network-assisted framework to build and update cybersecurity taxonomies and ontologies in cyber threat intelligence. Kougioumtzidou et al. ~\cite{Kougioumtzidou_2024} constructed the proposed taxonomy by identifying relevant entities via Natural Language Processing (NLP) techniques and regular expressions from a custom dataset. The resulting entities were further distilled via Term Extraction and Relation Extraction, piping into a Python library to construct the ontology. Finally, the taxonomies and ontologies were updated through semantic searches. While it was a refreshing take on the aspect of feature extraction and relationship correlation, the proposed framework lacked further ontology validation work, such as ABox reasoning. Moreover, since the custom datasets did not contain information about SCI, the scope of the approach was constrained compared to our work.

\SP aims to address the gaps identified by developing a comprehensive ontology that could be abstracted and express common constructs across the different branches of cybersecurity and to provide a unified standard for representing collected evidence aligned with the cybersecurity industry's widely used formats.

\subsection{Data Representation in Cybersecurity}
Data representation is crucial in cybersecurity, where collaborative investigation is paramount among blue teams. Additionally, a standardized format helps to streamline the analysis of data between various tools. However, due to this lack of standardization, existing tools and forensic frameworks do not integrate well with each other.

Additionally, numerous new smart devices need to be standardized due to the burgeoning field of IoT and the inception of smart cities. Stoyanova et al.~\cite{8950109} highlight that the origins of IoT data are often unverified, and the data itself typically lacks metadata, making it challenging to identify potential evidence. Furthermore, these IoT nodes are often in continuous operation and generate large amounts of data; if the data is dirty and unusable, it is just noise. Garfinkel's work on digital forensics XML~\cite{GARFINKEL2012161} was aimed at creating a standardized format to ensure interoperability among different forensics tools; however, with no digital forensics XML standard and the lack of fixed schema, there needed to be more adoption. Another significant contribution to the standardization of digital forensics is the Cyber-investigation Analysis Standard Expression (CASE) framework~\cite{CASEY201714}, developed by an international community of forensic practitioners and researchers. It provides a detailed and structured method of representing information from cyber incidents. While comprehensive, many organizations and companies may have proprietary environments incompatible with CASE~\cite{CASEY201714}. The Digital Forensics framework by Baguelin et al.~\cite{baguelin2010dff} was another noteworthy attempt at standardizing digital forensics, collecting multiple accolades and publications during its popularity. It was open source and supported on both Linux and Windows. However, user and developer interest shifted away, resulting in its gradual demise.

In the realm of cybersecurity countermeasures, Sánchez-García et al.~\cite{Sanchez_Garcia_2023} conducted a systematic mapping review of countermeasures and their taxonomies for risk treatment. Their work analyzes 26 taxonomies and catalogs of cybersecurity countermeasures, revealing trends, gaps, and the evolving focus of risk treatment strategies. Notable frameworks such as ISO27002 and NIST SP800-53 are highlighted for their widespread application. However, Sánchez-García et al. identified gaps in addressing risks associated with rapidly evolving domains, such as IoT and advanced persistent threats~\cite{Sanchez_Garcia_2023}. The study stresses the importance of holistic, standardized approaches integrating residual risk assessments while promoting interoperability across diverse systems. Despite its depth, the review broadly categorizes existing frameworks without proposing concrete mechanisms for their alignment or adaptation to emerging technologies.

\SP seeks to address these lapses by building on top of existing frameworks such as UCO/CASE ontology to ensure that it is interoperable and follows the standardized formats of ontology while staying technology agnostic to withstand the march of development.

\subsection{Cybersecurity Ontologies}
Ontologies are commonly used in academia, where formal naming and definitions are essential to establish the same baseline knowledge for research; its usefulness in the standardization of expertise has not been unnoticed by cybersecurity professionals. One of the most notable cybersecurity ontologies is UCO (Unified Cyber Ontology)~\cite{syed2016uco}, which provides a comprehensive list of definitions and support for various scenarios such as cybercrime. Another popular ontology would be the D3FEND Digital Artifact Ontology (DAO) from MITRE~\cite{kaloroumakis2024d3fend}, developed as part of its STIX framework; it is adept at facilitating the sharing of threat intelligence and supports widespread integration as part of STIX. IEEE also has its standard core ontologies, with a Common Core Cyber Ontology (C30) developed by its cyber ontology working group~\cite{cox2022cyberontology}. IEEE's C30 is designed as an overarching ontology covering aspects of cyberspace, serving as a domain-level representation of the various activities of cybersecurity.

Kahvedžić and Kechadi~\cite{Kahvedzic_2009} introduced the DIALOG framework, which offers a structured methodology for modeling, analyzing, and reusing digital forensic knowledge. By addressing the need for a comprehensive and reusable forensic knowledge repository, DIALOGUE reduces investigative process redundancy while enabling systematic, knowledge-driven investigations~\cite{Kahvedzic_2009}. Its modular design and scalability are particularly effective for addressing contemporary challenges, such as integrating forensic workflows across diverse technological and organizational environments. While DIALOG provides a robust foundation, the applicability to emerging fields like IoT and distributed environments such as SCI remains underexplored, limiting its utility in addressing modern digital forensic complexities.

With various organizations developing their cybersecurity ontologies, a few gaps have not been addressed during their development. A key issue is the inability to address new and upcoming cybersecurity developments, such as UCO providing a general, well-rounded cybersecurity ontology that can represent many scenarios; however, they lack the newer terms and definitions due to IoT and SCI developments. Another critical gap is the over specialization of ontologies. The MITRE DAO is highly adept at threat intelligence sharing but does not have a comprehensive vocabulary that could represent the myriad of possible scenarios. Widespread adoption is also critical for an ontology to be successful. With a lack of implementation in cybersecurity tools, IEEE's C30 faces issues with industry usage since it has not been used with any of the typical cybersecurity applications, unlike UCO or DAO, which has seen usage in tools such as Autopsy~\cite{Autopsy}. 

\SP extends from UCO, ensuring that it is interoperable and can also be used with tools that already incorporate UCO. \SP also extends the terminology to cover emerging technologies, thus providing the representation of cutting-edge cybercrime and investigation scenarios. By integrating these advances, \SP presents a forward-looking solution that addresses the unique challenges posed by emerging cybercrime scenarios and technologies.
\section{Discussion and Conclusion} 
\label{Conclusion}

% Discuss future prospects
With the danger of cybercrime in the constantly developing field of technology, it is paramount that the LEA and DFI are suitably equipped to deal with this burgeoning issue. Additionally with the advent of smart cities with technology interwoven at every layer of the system, communicating and sharing key vital information is critical to defeating these adversaries.

We proposed \SP, a modular expansion profile of the Unified Cyber Ontology (UCO) and the Cyber-investigation Analysis Standard Expression (CASE) ontology. Noting that building a SCI ontology just by relying on UCO and CASE could be a laborious and semantically challenging endeavor, we proposed \SP as an expansion. \SP adheres to the original design requirements of UCO and CASE, and also achieves tool interoperability and the ability for collaborative investigations.

\SP is technology-agnostic while also adhering to international standards such as the ISO standards. Additionally, it contains enough granularity to allow users to pinpoint key information while also ensuring it is able to capture abstract definitions that can cover emerging technologies. For extensibility and further research 
in cybercrime and digital forensics, we have made the entire \SP ontology available in the following URL:
\begin{center}
\url{https://ontology.scopeontology.org}
\end{center}

\section{Acknowledgments}

This research is partially supported by Infocomm Media Development Authority under its Future Communications Research \& Development Program (Award number FCP-SUTD-RG-2022-017) and National Research Foundation, Singapore, under its National Satellite of Excellence Programme “Design Science and Technology for Secure Critical Infrastructure: Phase II” (Award No: NRF-NCR25-NSOE05-0001). Any opinions, findings and conclusions or recommendations expressed in this material are those of the author(s) and do not reflect the views of the respective funding agencies.

{%\small 
	\bibliographystyle{unsrt}	
	\bibliography{references}
}

% \end{thebibliography}
% \vspace{12pt}
% \color{red}
% IEEE conference templates contain guidance text for composing and formatting conference papers. Please ensure that all template text is removed from your conference paper prior to submission to the conference. Failure to remove the template text from your paper may result in your paper not being published.

\end{document}